\documentclass[conference]{IEEEtran}
\usepackage{graphics,float}
\graphicspath{{figs/}}
\usepackage[linesnumbered,algoruled,boxed,lined]{algorithm2e}
\let\labelindent\relax
\usepackage{enumitem}
\usepackage{caption}
\usepackage{pifont}
\usepackage{amsmath,amsfonts}
\usepackage{xspace}
\usepackage{tikz}
\usepackage{gensymb}
\usepackage{array}
\usepackage{bm}
\usepackage{subfigure}
\usepackage{multirow}
\usepackage{graphicx,tabularx,ragged2e,booktabs}
\usepackage[flushleft]{threeparttable}
\usepackage{cite}
\usepackage{url}

\ifCLASSINFOpdf
\else
\fi
\newcommand*\circled[1]{\tikz[baseline=(char.base)]{
            \node[shape=circle,draw,inner sep=0.5pt] (char) {#1};}}
\newcommand{\alias}{{\sc Vrifle}\xspace}
\newcommand{\xmark}{\ding{55}}%
\newcommand{\blue}[1]{\textcolor{black}{#1}}
\DeclareCaptionFont{mycaption}{\fontsize{9}{10}\selectfont}
\begin{document}
%
\title{Inaudible Adversarial Perturbation: Manipulating the Recognition of User Speech in Real Time}
\IEEEoverridecommandlockouts
\makeatletter\def\@IEEEpubidpullup{6.5\baselineskip}\makeatother
\IEEEpubid{\parbox{\columnwidth}{
    Network and Distributed System Security (NDSS) Symposium 2024\\
    26 February - 1 March 2024, San Diego, CA, USA\\
    ISBN 1-891562-93-2\\
    https://dx.doi.org/10.14722/ndss.2024.23030\\
    www.ndss-symposium.org
}
\hspace{\columnsep}\makebox[\columnwidth]{}}
\author{\IEEEauthorblockN{Xinfeng Li}
	\IEEEauthorblockA{Zhejiang University\\
		xinfengli@zju.edu.cn}\\
	\IEEEauthorblockN{Zihan Zeng}
	\IEEEauthorblockA{Zhejiang University\\
		zengzh@zju.edu.cn}
	\and
        \IEEEauthorblockN{Chen Yan\IEEEauthorrefmark{2}\thanks{\dag Chen Yan and Xiaoyu Ji are the corresponding authors}}
	\IEEEauthorblockA{Zhejiang University\\
		yanchen@zju.edu.cn}\\
        \IEEEauthorblockN{Xiaoyu Ji\IEEEauthorrefmark{2}}
	\IEEEauthorblockA{Zhejiang University\\
		xji@zju.edu.cn}
	\and
        \IEEEauthorblockN{Xuancun Lu}
	\IEEEauthorblockA{Zhejiang University\\
		xuancun\_lu@zju.edu.cn}\\
	\IEEEauthorblockN{{Wenyuan Xu}}
	\IEEEauthorblockA{Zhejiang University\\
		wyxu@zju.edu.cn}	
}

\maketitle


\begin{abstract}

Automatic speech recognition (ASR) systems have been shown to be vulnerable to adversarial examples (AEs). Recent success all assumes that users will not notice or disrupt the attack process despite the existence of music/noise-like sounds and spontaneous responses from voice assistants. Nonetheless, in practical user-present scenarios, user awareness may nullify existing attack attempts that launch unexpected sounds or ASR usage. 
In this paper, we seek to bridge the gap in existing research and extend the attack to user-present scenarios. We propose \alias, an inaudible adversarial perturbation (IAP) attack via ultrasound delivery that can manipulate ASRs as a user speaks.
The inherent differences between audible sounds and ultrasounds make IAP delivery face unprecedented challenges such as distortion, noise, and instability. In this regard, we design a novel ultrasonic transformation model to enhance the crafted perturbation to be physically effective and even survive long-distance delivery.
We further enable \alias's robustness by adopting a series of augmentation on user and real-world variations during the generation process.
In this way, \alias features an effective real-time manipulation of the ASR output from different distances and under any speech of users, with an \textit{alter-and-mute} strategy that suppresses the impact of user disruption.
Our extensive experiments in both digital and physical worlds verify \alias's effectiveness under various configurations, robustness against six kinds of defenses, and universality in a targeted manner. We also show that \alias can be delivered with a portable attack device and even everyday-life loudspeakers.
\end{abstract}



\section{Introduction}
Automatic speech recognition (ASR) enables computers to transcribe human speech and is essential in a wide range of voice applications such as voice assistants (VAs) and audio transcription APIs~\cite{azureasr,googleasr}.
Prior studies have shown that ASR models are vulnerable to adversarial examples (AEs) that sound benign to humans but are recognized incorrectly by models. 
As stealthiness is a basic requirement for AEs, existing works largely focus on reducing the audibility of AEs so that they might not cause human suspicion when being heard~\cite{schonherr2018adversarial,qin2019imperceptible}. In addition, the class of inaudible attacks~\cite{zhang2017dolphinattack,sugawara2020light} avoids being perceived by human ears using high-frequency ultrasound/laser.
However, few of them have considered attacks in user-present scenarios, where users may notice unexpected events of the ASR service and can mitigate the attack's consequence.
For instance, though AEs and inaudible attacks may not sound suspicious, a voice assistant will always provide feedback (e.g., vocal prompt or LED blinking) after receiving voice commands. Alert users may still notice the false wake-up or abnormal feedback caused by an attack and speak remedy commands to correct the mistake, limiting the attack's impact in real life.

\begin{figure}
    \centering
    \includegraphics[width=0.48\textwidth]{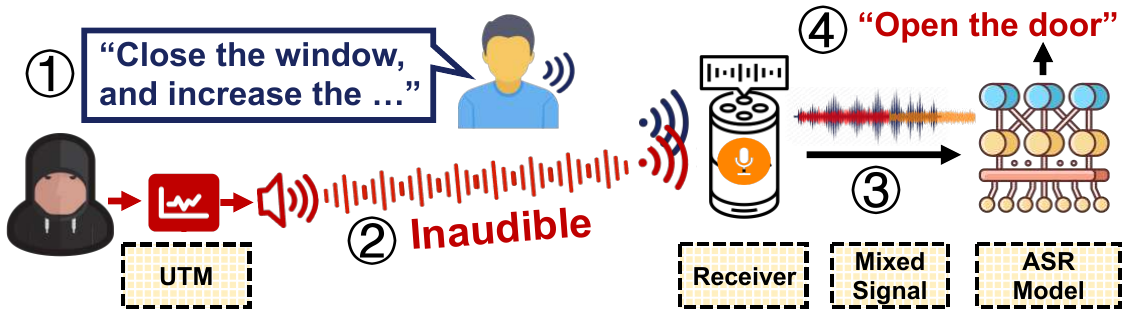}
    \caption{\ding{172}When a user uses the ASR service, \ding{173}an adversary injects inaudible adversarial perturbations crafted based on the ultrasonic transformation model (UTM) into a receiver. \ding{174}The mixed signal of the user command (blue) and demodulated perturbations (red \& yellow) can \ding{175}fool the ASR model into the adversary-desired intent.}
    \label{fig:fig1}
    \vspace{-15pt}
\end{figure}

In this paper, we aim to propose \alias\footnote{Demo: https://sites.google.com/view/Vrifle}---an \textit{inaudible adversarial perturbation} (IAP) attack that can extend to this scenario. Its basic idea is to inject IAPs while the user speaks to the ASR service and alter the recognition result in real time, as shown in Fig.~\ref{fig:fig1}. Since the voice assistant itself is responding to user commands (e.g., LED blinking), tampering with the user's speech is less noticeable at this time. 
\blue{
But such a user-present scenario also imposes higher requirements on attack stealthiness because users are more sensitive to environmental sounds while using ASR services. 
Moreover, given that adversaries have no prior knowledge of user's speech content and timing, this critical scenario necessitates that \alias exhibits a high level of universality to guarantee the achievement of the adversary-desired intent in any context.
Therefore, we envision \alias as a truly inaudible and robust framework for real-time IAPs delivery, which can also address variable user factors, such as speech content, vocalization time, speech volume, and environmental conditions, while remaining physically effective even at long distances or using portable/everyday-life devices.
Overall, materializing \alias that attains the above goals is challenging in three aspects.}

\begin{itemize}[leftmargin=*]
\item \textit{How to achieve adversarial perturbations that are \textbf{universal} while \textbf{completely inaudible} to \textbf{user auditory}?}
\end{itemize}

The trade-off between universality and stealthiness has been a long-standing challenge in audio AE attacks. Almost all previous works have prioritized stealthiness and introduced imperceptibility constraints during optimization, such as $\epsilon$ and L2-Norm~\cite{carlini2018audio}, or by adjusting audio forms, e.g., designing it as short pulses~\cite{guo2022specpatch}. Nonetheless, this greatly compromises the universality of adversarial perturbations and they are still audible and can be heard when users are nearby.
We seek to implement an inaudible adversarial perturbation beyond the human auditory range (20~Hz$\sim$20~kHz) in an ultrasound-based attack manner~\cite{zhang2017dolphinattack},
which can make microphones receive our IAP by exploiting their inherent nonlinearity vulnerability. As such, IAPs are no longer limited by stealthiness constraints, holding a vast optimization space with more feasible solutions. 
Unlike the audible-band perturbations devised to be short to mitigate user auditory, our IAPs enable the adversary to significantly increase their length, which further expands the optimization scope and facilities highly universal attacks.

\begin{itemize}[leftmargin=*]
\item \textit{How to alter the recognition of user speech in real time despite the presence of \textbf{user disruption}?}
\end{itemize}

Although we have bypassed \textit{user auditory}, realizing such an attack against ASR in real time faces a few more challenges. 
\blue{
\textit{User disruption} cannot be ignored in this scenario, which includes: \ding{202} The user's speech can disrupt the intent of IAPs when both audio signals are superimposed. While universal AEs~\cite{li2020advpulse,guo2022specpatch} are shown to resist this case, our preliminary investigation validates that direct ultrasound-based attacks will fail due to such interference. \ding{203} User commands can be much longer (e.g., 5s) than 0.5s audible-band short perturbations that affect only a few input frames, thus the exceeded user instructions will impact the entire ASR transcription. \ding{204} Users may notice that malicious behavior being executed and therefore block the attack by issuing remedy commands. 
In addition, there are user-induced factors that make \textit{user disruption} more complex and can compromise IAPs' effectiveness, including unpredictable content and timing of user speech, as well as the influence of the user's environment and speaking habits on speech reverberation and loudness.}

To address these issues, we augment the optimization process of IAPs by using multiple speech clips in public corpus, introducing randomness within the preset time range, as well as considering the various user's speech loudness and reverberation. Thereby, \alias can be applied in a content-agnostic, synchronization-aided, user factors-robust manner. Moreover, we overcome \textit{user disruption} by materializing both silence and universal perturbations in the targeted manner to ensure the arbitrary utterance length cannot pose impacts on adversary-desired intent, without requiring any knowledge.
Based on the above design, adversaries can present two more hidden attack strategies, involving \textit{No-feedback Attack} and \textit{Man-in-the-middle Attack} in the threat model.

\begin{table}\centering
    \small 
    \renewcommand{\arraystretch}{0.5}
    \renewcommand\tabcolsep{5.5pt}
    \begin{threeparttable}[t]
    \setlength{\abovecaptionskip}{5pt}%
    \setlength{\belowcaptionskip}{0pt}%
    
    \caption{Compared with existing works}
    \begin{tabular}{@{}lcccc@{}}
        \toprule
        \textbf{Method} & \textbf{Constraint$^\ddagger$} & \textbf{Auditory$^\natural$}  & \textbf{Disruption$^\star$} 	& \textbf{Dist.$^\dagger$}\\ \midrule
        Carlini.~\cite{carlini2016hidden}  &   $\relbar$   &  Noise  & \xmark    & 1.5m      \\ \midrule
        Abdullah~\cite{abdullah2019ndss} &   $\relbar$   &  Noise  & \xmark    & 0.3m      \\ \midrule
        CW~\cite{carlini2018audio}  &  L2-norm, $\epsilon$   &  Speech  & \xmark    & \xmark       \\ \midrule
        Schönherr~\cite{schonherr2018adversarial} &  Psyc.   &  Speech  & \xmark    & \xmark       \\ \midrule
        Comman.~\cite{yuan2018commandersong} &  $\epsilon$   &  Song  & \xmark    & 1.5m  \\ \midrule
        Qin.~\cite{qin2019imperceptible}  &  Psyc.   &  Speech  & \xmark    & $\relbar$  \\ \midrule
        Meta-Qual~\cite{chen2020metamorph}  &  L2-norm, $\epsilon$   &  Song  & \xmark    & 4m  \\ \midrule
        FakeBob~\cite{chen2021fakebob} &  $\epsilon$   &  Speech  & \xmark    & 2m  \\ \midrule
        AdvPulse~\cite{li2020advpulse} &  L2-norm, $\epsilon$   &  Ambient  & \ding{119} & 2.7m  \\ \midrule
        SpecPatch~\cite{guo2022specpatch} &  L2-norm   &  Pulse  & \ding{119}  &  1m    \\ \midrule
        \textbf{Ours}   &  \textbf{None}   &  \textbf{Inaudible}  &  \ding{108}  & \textbf{10m}  \\ \bottomrule
        \end{tabular}
        \begin{tablenotes}[flushleft]
        \item[] \vspace{-2pt}\hspace{-2pt}\small 
        (i) $\ddagger$: The constraints used to guarantee imperceptibility during optimization. ``$-$'' means the method only considers incomprehensibility to humans. $\epsilon$ means limiting the absolute magnitude of perturbations with a constant $\epsilon$. $L_2$-Norm means adding an $L_2$-Norm term in the objective function. ``Psyc.'' means psychoacoustic hiding. ``None'' means no stealthiness constraints. 
        (ii) $\natural$: The objective \textit{\textbf{user auditory}} of AEs. Ambient means ambient sounds.
        (iii) $\star$: \ding{108}: fully tackles \textit{\textbf{user disruption}}. \ding{119}: tackles case \ding{202}. \xmark: fails by \textit{\textbf{user disruption}}.
        (iv) $\dagger$: \xmark: the attack is not physically available. $\relbar$: not reported.
        \end{tablenotes}
        \vspace{-10pt}
    \label{tab:compare}
    \end{threeparttable}
\end{table}

\begin{itemize}[leftmargin=*]
\item \textit{How to guarantee inaudible adversarial perturbations are \textbf{physically effective} after ultrasonic delivery?}
\end{itemize}

Though inaudible attacks have demonstrated voice command injection using ultrasound and laser~\cite{sugawara2020light}, \textit{it is unknown whether fine-grained IAPs can be delivered via such signals} as the ultrasound channel is reported to be lossy and distorted~\cite{li2023learning}. Thus, maintaining the effectiveness of IAPs after undergoing a series of modulation, transmission, and demodulation processes in the physical world is not trivial based on prior AEs~\cite{schonherr2020imperio}. Ultrasound is intrinsically distinct from audible sounds in the high-directional propagation and varying soundfield. Additionally, the nonlinear distortion, anomalous noises, and hardware-induced instability that are unique to ultrasound make existing acoustic channel modeling methods inapplicable.

To overcome the challenge, we make the first attempt to establish an ultrasonic transformation model, which consists of tackling variable ultrasound-induced anomalous noises, obtaining ultrasound frequency response (UFR), and enabling location-variable attacks. 
Based on this transformation, we can precisely estimate \alias's pattern of ultrasonic delivery during its optimization, thereby making it physically effective and survive long-distance delivery. Moreover, to enable more covert IAP attacks with portable devices and off-the-shelf loudspeakers, we implement a narrow bandwidth upper-sideband modulation (USB-AM) mechanism to ensure the attack range and inaudibility of \alias with simplified devices.


Tab.~\ref{tab:compare} compares \alias with several existing works. 
We conduct extensive experiments in both digital and physical worlds to evaluate \alias's effectiveness under various configurations (e.g., extend attack range to 10m) and robustness against six kinds of defenses. Our single silence IAP muting up to 27,531 unseen user utterances, likewise, universal IAP altering 18,956, proving \alias's universality. 
\blue{Our design also expands the attack methodology to more covert portable attack devices and everyday-life loudspeakers, enabling the \alias delivery in a stealthier form.} 
Our contribution can be summarized as follows:




\begin{itemize}[leftmargin=*]
    \item To the best of our knowledge, \alias is the first universal inaudible adversarial perturbation attack that can extend to scenarios when users use ASR services, revealing a new attack surface against ASR models. \alias is completely inaudible, holds vast optimization space, and enables long-range attacks (10m).
    \item We make the first attempt to establish an ultrasound transformation model, which overcomes the unique challenges in the ultrasound channel and precisely characterizes it, enabling our fine-grained IAPs delivery to be physically effective.
    \item We conduct extensive experiments under various configurations in the digital and physical world to validate the effectiveness, robustness, and universality of \alias, and validate the attack using portable/everyday-life devices.
    
\end{itemize}


\section{Background}
\subsection{Automatic Speech Recognition}\label{background_ASR}
Automatic speech recognition (ASR) systems, e.g., voice assistants, receive and recognize speech commands; then perform execution according to certain rules. 
Hidden Markov models (HMM)~\cite{baum1966statistical} and dynamic time warping (DTW)~\cite{berndt1994using} are two traditional statistical techniques for performing speech recognition.
With the development of deep learning, the end-to-end neural ASR models have gone mainstream, such as RNN-T~\cite{graves2012sequence} and DeepSpeech2~\cite{amodei2016deep}. 
A typical end-to-end ASR system pipeline includes four main components: \circled{1} \textit{Spectrum generator}: converts raw audio into spectrum features, e.g., Filter Bank (Fbank), Mel-Frequency Cepstral Coefficients (MFCC), etc. \circled{2} \textit{Neural acoustic model}: takes spectrums as input and outputs a matrix of probabilities over linguistic units (e.g., phoneme, syllable, or word) over time. For instance, English ASR is widely modeled with 29 basic units (also known as tokens), including characters a\textasciitilde z, space, apostrophe, and blank symbol $\phi$. \circled{3} \textit{Decoder}: generates possible sentences from the probability matrix, also optionally coupled with an n-gram language model to impose word-level constraints. The Connectionist Temporal Classification (CTC) module is a typical decoder that sums over all possible alignments that may reduce to the same token sequence, whereby ``o$\phi$kk$\phi$a$\phi$y'' and ``o$\phi$k$\phi$aa$\phi$yy'' are regarded as the same ``okay''. \circled{4} \textit{Punctuation and capitalization model}: formats the generated text for easier human consumption.

\subsection{Audio Adversarial Examples}
Adversarial examples (AEs)~\cite{carlini2018audio,schonherr2018adversarial,yuan2018commandersong,chen2020metamorph} use specialised inputs created with the purpose of confusing a neural network, resulting in the misclassification of a given input. In the audio domain, by adding a crafted perturbation $\delta$ with some constraints $\epsilon$ throughout the original benign audio $x$, the ASR model will be fooled to transcribe a perturbed speech into the targeted text $y_t$, e.g., ``take the picture''. To craft an adversarial example, an adversary may leverage the optimization function:
\begin{equation}
\begin{aligned}
    & {minimize}~ \mathcal{L}(f(x+\delta), y_t) + \alpha\cdot\Vert{\delta}\Vert_{p}\\
    & {s.t.}~ \delta \in [-\epsilon, \epsilon]^n, (\epsilon\textless 0.01)
\end{aligned}
\label{equ:audible_ae_formula}
\end{equation}
where the ASR functions as $f(\cdot)$ that takes an input waveform and outputs the probability matrix. $\mathcal{L}(f(\cdot),y_t)$ is the CTC loss function denoting the distance between the model output of the adversarial example and the target. $\|\cdot\|_{p}$ means the $L_p$ norm. $\alpha$ is a penalty term to limit the $L_p$. $\epsilon$ denotes the upper bound of the perturbation. 
Recently, the concept of universal adversarial perturbation is proposed, making AEs valid regardless of the user commands.
To make the AEs more concealed, the creating approaches are extended to psychoacoustic hiding~\cite{schonherr2018adversarial,qin2019imperceptible} and shorter pulses~\cite{li2020advpulse}. However, existing efforts cannot fundamentally avoid being perceived by human ears.

\begin{figure}[t]
	\centering
	\includegraphics[width=0.4\textwidth]{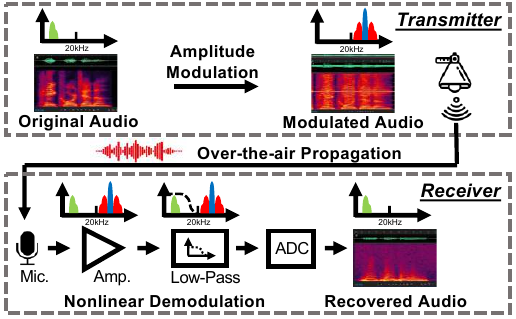}
	\caption{\label{fig:dolphinattack_pipeline}Diagram of inaudible attacks (carrier: blue, baseband: green).}
        \addvspace{-15pt}
\end{figure}

\subsection{Ultrasound-based Attacks}\label{background_vulnerability_VCS}
Inaudible attacks modulate the audio baseband on high-frequency carriers to the inaudible band of human ears ($>$20~kHz) and exploit microphones' nonlinear vulnerability, so that ASRs can receive the malicious audio while humans cannot perceive it.
Recently, inaudible attacks have been extended from ultrasonic carrier~\cite{zhang2017dolphinattack,roy2018inaudible} to various forms, such as solid conduction~\cite{yan2020surfingattack}, laser~\cite{sugawara2020light}, capacitor~\cite{zhang2021capspeaker}, power line~\cite{wang2022ghosttalk}, etc., forming a class of highly threatening and comprehensive covert attacks. 
We take the representative ultrasound-based attack~\cite{zhang2017dolphinattack} to present the principle of inaudible attacks shown in Fig.~\ref{fig:dolphinattack_pipeline}.
First, the original audio is double-sideband (DSB) modulated on an ultrasound carrier via amplitude modulation (AM). Second, the DSB-AM audio is emitted from the ultrasonic transducer and propagates over the air. Third, after the microphone receives the signal, audio modulated on the high-frequency carrier will be recovered into the audible band \textit{before the low-pass filters and ADC} due to nonlinear effects of the microphone's diaphragm and amplifier. Thus, though the ultrasound carrier is finally filtered, the demodulated audio still survives and functions to ASR.
The nonlinear demodulation is formulated as follows: 
\begin{equation}\label{equ:nonlinearity}
    S_{out}(t) = \sum_{i=1}^{\infty}k_{i}s^{i}(t)=k_1 s(t) + k_2 s^2(t) + k_3 s^3(t) + ...
\end{equation}
where $s(t)$ and $S_{out}(t)$ indicate the input AM signal and amplifier's output, respectively. 
The even order terms, i.e., $k_2$, $k_4$ are the key in recovering the original audio~\cite{roy2017backdoor}.
Notably, such an ultrasound channel is lossy as the recovered audio samples differ from original ones. Our investigation demonstrates that the channel is also challenging to model~\ref{sec:ultrasound_observation}.
\subsection{Threat Model}\label{sec:threat_model}
\textbf{Attack Scenarios:}\label{attack_scenario}
We consider attacks in user-present scenarios where the user may notice unintended events of ASR services. Such scenarios involve two entities:\\
\underline{\textbf{Victim}:} The victim user is alert to any strange sounds (e.g., noise, music, pulses) within human auditory. The user can speak an arbitrary command to the smart speaker. Once the user notices attacks, he/she can speak a remedy command to the smart speaker.\\
\underline{\textbf{Adversary}:} The adversary prepares IAPs for specific intentions offline and alters user command in real time by delivering IAPs \ding{172}at a distance from the victim with an ultrasonic transmitter through the window, \ding{173}physically close with a handheld portable device, or \ding{174}with a preset off-the-shelf loudspeaker. The adversary's goals are providing wrong information to intelligent voice customer service, compromising VAs to execute malicious commands or be in denial-of-service mode, etc. The adversary can attack more covertly with two strategies: \textit{1) No-feedback Attack}: Prevent the user from hearing VA's vocal prompt by ``Mute volume and turn off the WiFi''. \textit{2) Man-in-the-middle Attack}: Once the user's intent is satisfied, the attack may be much less suspicious, i.e., while delivering the adversarial perturbation and alter user commands, adversaries can record the user commands and then replay it by traditional ultrasound-based attack means. 

\textbf{Attacker Capability:}
Distinct from the previous works~\cite{carlini2018audio,chen2020metamorph,yuan2018commandersong,qin2019imperceptible,chen2021fakebob,schonherr2018adversarial,li2023tuner} that require the user's speech samples in advance to craft adversarial perturbation, we assume the adversaries have no knowledge of what the user will speak during performing attacks.
In line with the widely adopted settings in prior works~\cite{yuan2018commandersong,carlini2018audio,chen2020metamorph,qin2019imperceptible,guo2022specpatch}, we assume the attackers have prior knowledge of the target ASR model for obtaining the gradient information during optimization.
The adversaries have access to the user's recording device, e.g., borrow a smartphone of the same brand, based on which the adversary can model the ultrasonic transformation, and then create the IAP in advance.
We assume adversaries have the flexibility to deploy the hidden ultrasonic transmitter nearly or at a distance, and the recording device is in its line of sight. Additionally, adversaries can also utilize stealthy portable devices and off-the-shelf loudspeakers in everyday-life scenarios to deliver \alias.

\begin{figure}
    \centering
    \includegraphics[width=0.45\textwidth]{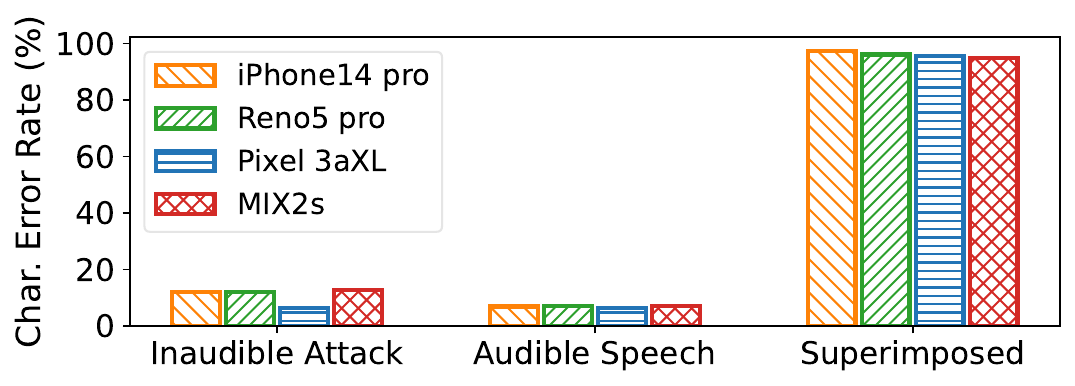}
    \caption{CER of four recording devices under three settings. }
    \label{fig:preliminary_wer}
    \vspace{-10pt}
\end{figure}

\section{Preliminary Investigation}\label{sec:preliminary}
\subsection{\blue{Failure of Traditional Inaudible Attacks}}\label{pre_inaudible_attack}
\blue{
Given the purpose of avoiding alerting users, directly injecting malicious commands into ASR systems using laser-~\cite{sugawara2020light} or ultrasound-based~\cite{zhang2017dolphinattack,roy2018inaudible} inaudible attacks is intuitive. Although laser-based attacks can reach an 100m attack range, we choose ultrasound instead of laser for three practical reasons: (1) The laser spot on the microphone is visible and will alert users immediately; (2) The laser-based attack requires strict line-of-sight alignment; and (3) The severe channel distortion of laser-delivered attacks may nullify fine-grained adversarial perturbations.}

\blue{
To examine whether traditional ultrasound-based attacks can manipulate ASRs into recognizing the modulated malicious commands while users are speaking, we need to ensure that the ultrasonic carrier frequency is optimal. Therefore, we first employ an ultrasonic Vifa~\cite{vifa} to launch a wide-range carrier sweeping from 20$\sim$40~kHz. By analyzing the signal-to-noise ratio (SNR) of demodulated basebands, we justify the optimal frequency of four recording devices, i.e., iPhone14 pro: 24.7~kHz, Reno5 pro: 27.7~kHz, Pixel 3aXL: 25.6~kHz, and MIX2s: 25.1~kHz, respectively. This result is consistent with DolphinAttack~\cite{zhang2017dolphinattack}, which reveals most devices' optimal attack carrier frequency is around 25~kHz (22.6$\sim$27.9~kHz). In this way, we set the default carrier frequency to 25~kHz, whose advantages are two-fold: (1) Due to lower airborne attenuation, 25~kHz also benefits longer-range attacks than high-frequency carriers (e.g., 40kHz); (2) Moreover, 25~kHz as one of the most typical parameters for commercial ultrasonic transducers that cost as low as 0.14\$ per unit~\cite{25kHz_ultra_transducer}, making the attack cost-effective.
}

\blue{
Although the optimal attack frequency is determined, traditional ultrasound-based attacks still fail due to \textit{user disruption}. Specifically, we select 10 text-to-speech commands listed in Tab.~\ref{tab:diff_commands} (e.g., ``turn on airplane mode'') as the basebands. 
Four smartphones 50~cm away serve as recording devices that recover the AM signal into audible-band speech. 
For benign command samples, we randomly select 20 utterances from the popular fluent speech commands dataset~\cite{fluent2020commands} to be played via a loudspeaker and recorded by identical smartphones.
We also perform simultaneous emissions of both signals, so they are superimposed on each other. 
For each recording device, we collected $10\times20=200$ mixed samples and calculated each sample's character error rate (CER) through the Azure speech-to-text API~\cite{azureasr}. As shown in Fig.~\ref{fig:preliminary_wer}, the direct ultrasound-based attacks and benign audio are well recognized by ASR models, with average CER of 10.8\% and 6.88\%, respectively. Nevertheless, once attack emission and user's voice coincide, the attack performance (i.e., 10 malicious commands as the target transcription) will severely degrade to an average CER up to 96.01\%, even if we have boosted its power\footnote{To facilitate ultrasound-based attacks, we set the volume up to 95 dB, and that of audible benign speech is 70$\sim$75 dB.}. We believe it is a consequence that when ASRs process the mixed samples, each sampling point of the malicious signal sequence is affected by the human voice, making the acoustic features extracted by the ASR deviate from adversaries' anticipation.}




\subsection{Ultrasonic Adversarial Perturbation Delivery}\label{sec:preliminary_uap_delivery}
We envision that the above failure can be addressed by leveraging the vulnerability of ASR models to craft universal adversarial perturbations. Notably, it is promising to deliver the perturbation in an ultrasound-based manner to eventually reach the goal, i.e., the adversary can alter any user commands into a targeted one while guaranteeing entirely inaudible to the victim. However, we find the well-trained perturbations that are effective in digital domain all fail after being directly modulated and emitted by the ultrasound-based attack method (results are also given in \textsection\ref{sec:eval_utm}, \textit{G2}). 

Since the ultrasonic channel is lossy and distorted, to obtain a perturbation that can still effectively tamper with user commands after a series of processing based on ultrasound-based attack mechanisms and over-the-air delivery, i.e., the pipeline shown in Fig.~\ref{fig:dolphinattack_pipeline}, we need to precisely model the transformation from a perturbation in the digital domain to its physical version.
However, generalizing an AE from the digital to the physical world is inherently difficult, which has been proved by substantial research in both computer vision and audio community~\cite{jan2019connecting,carlini2018audio,chen2020metamorph,schonherr2020imperio,li2020advpulse,deng2022fencesitter,guo2022specpatch}. 
This issue in the audio domain refers to the fact that played-out speech samples are subject to signal distortion and environment interference (i.e., reverberation, attenuation, and noises). 
Previous audible-band works~\cite{rirIJCAI,schonherr2020imperio} have paid efforts in simulating the physical world by adopting room impulse response (RIR) during the AE optimization process to close the gap between the digital and physical world. Moreover, no work has yet been proposed on modeling ultrasonic delivery.
We are motivated to investigate the feasibility of applying audible-band modeling technologies to our unique ultrasonic case.
\begin{figure}[t]
    \centering
    \includegraphics[width=0.35\textwidth]{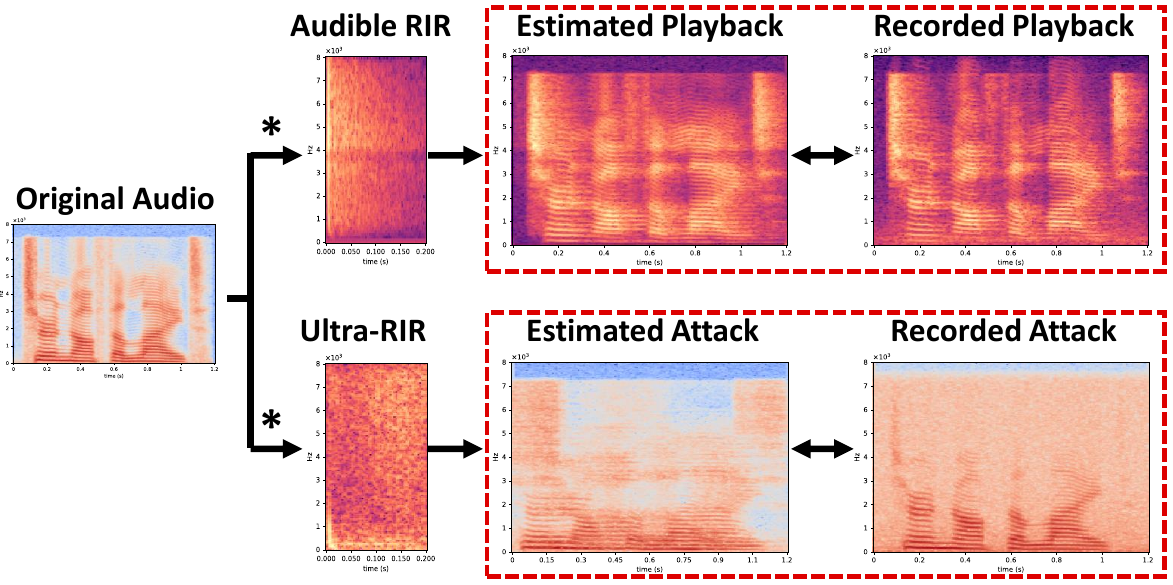}
    \caption{\label{fig:est_rir}Comparison between the audible RIR and ultra-RIR in estimating the digital-to-physical transformation.}
    \vspace{-10pt}
\end{figure}
\subsection{Attempts at Ultrasound Delivery Modeling}
In this subsection, we elaborate on applying two potentially feasible modeling methods for our case.


\subsubsection{Modeled by room impulse response}\label{sec:classic_rir}
Inspired by the success of audible-band AEs~\cite{rirIJCAI,schonherr2020imperio} drawing on the ability of RIR, which describes the reverberation and attenuation during audible sound propagation, we envisage that a similar RIR idea can generalize to characterize the ultrasound transformation process. Specifically, they exploit existing RIR databases~\cite{jeub2009binaural,nakamura2000acoustical} by convolving random RIR clips with digital adversarial signals in the optimization process, simulating the audio recorded by the receiver in various scenes, e.g., large concert hall and narrow corridor.
Therefore, we modulate the ideal impulse signal as the baseband on an ultrasound carrier and receive it on the recording device. With the obtained ``ultra-RIR'', we perform convolution with the original audio, whose output are expected to well represent the actually demodulated inaudible attack's result. As a comparison, we also conduct similar operations via a JBL loudspeaker for audible audio.
Fig.~\ref{fig:est_rir} shows that the estimated audible audio with RIR is very close to the actual playback. However, for the inaudible aspect, there is significant gap between the recorded attack audio and the estimated using ultra-RIR.
We believe the reason for such a mismatch is that the RIR rationale relies on the linear time-invariant (LTI) system prerequisite. However, the transformation is nonlinear because ultrasound-based attacks leverage microphones' nonlinearity vulnerability. 


\begin{figure}[t]
    \centering
    \includegraphics[width=0.3\textwidth]{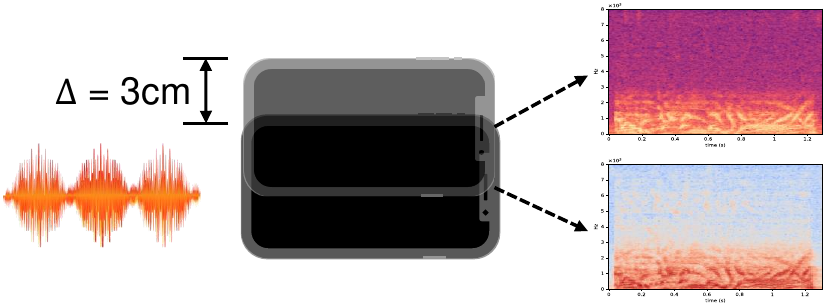}
    \caption{\label{fig:AE_sensitivity}Illustration of displacement-induced changes in recorded audio.}
    \vspace{-10pt}
\end{figure}

\subsubsection{Modeled by Neural Network}\label{sec:classic_nn}
Since RIR is originally designed for LTI systems, neural networks with excellent nonlinear fitting capability should work well given their success in various tasks, e.g., speech denosing~\cite{hu2020dccrn} and image printing distortion~\cite{jan2019connecting}. 
Considering that an adversary expects a practical transformation model with minimal effort (i.e., dataset requirements) while guaranteeing its generality, we implement a multi-layer perception model (MLP) with only 60k parameters, using 120-second aligned original and ultrasound-based attack audio pairs. We find that the MLP can achieve a generalized capability of mapping digital-to-physical world spectrums between unseen pairs, but with position-dependent constraints. As shown in Fig.~\ref{fig:AE_sensitivity}, a slight position displacement (3~cm) leads to an apparent change (i.e. bringing anomalous noises) in the recovered baseband, which can cause the trained network to fail to estimate the recorded audio at various positions. 
Overall, although MLP builds a functional mapping for the nonlinear ultrasound transformation in a fixed relative position, it is too restricted due to the nature of ultrasound (cf. \textsection\ref{sec:ultrasound_observation}). Besides, adopting distance $d$ and angle $\theta$ as conditional network parameters might help, but collecting data for each position is endless.

\begin{figure}[t]
    \centering
    \includegraphics[width=0.36\textwidth]{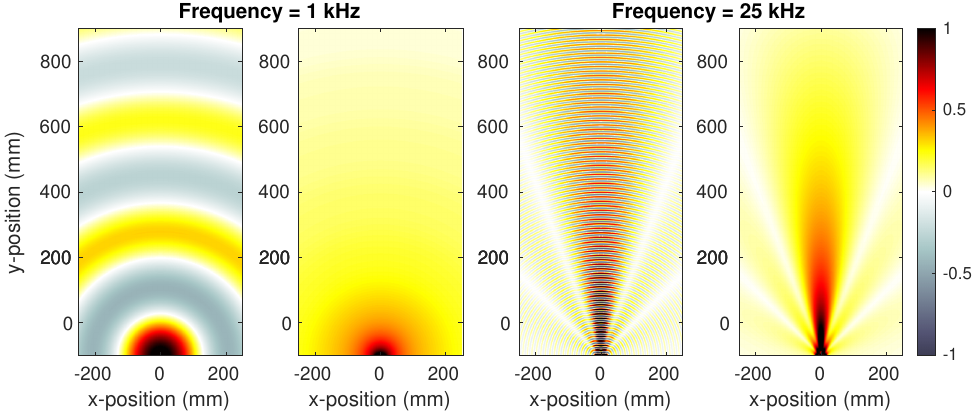}
    \caption{\label{fig:soundfield}2-D sound fields simulation for comparing the audible wave with ultrasound.}
    \vspace{-15pt}
\end{figure}

\begin{figure*}[!t]
    \centering
    \includegraphics[width=0.95\textwidth]{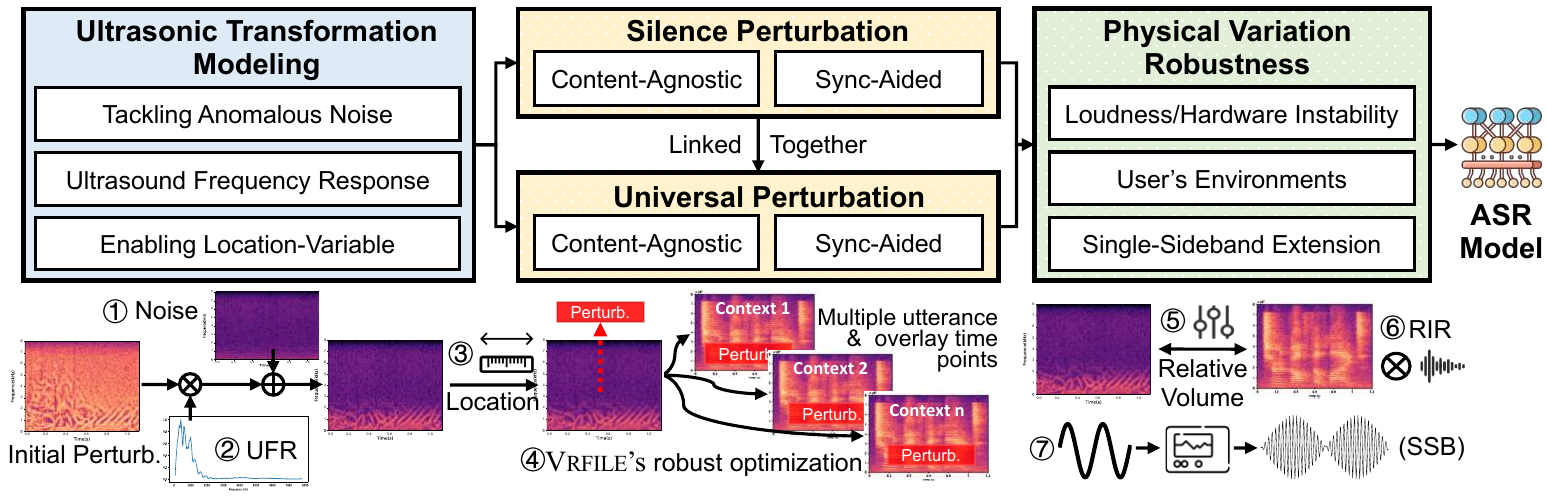}
    \caption{Workflow of \alias. \ding{172}-\ding{174}: the ultrasonic transformation precisely describes the perturbation changes during physical delivery. \ding{175}: the transformed perturbation is involved into optimization for silence and universal attack purpose. \ding{176}-\ding{178}: we boost the attack's physical-world robustness from multiple aspects.}
    \label{fig:design_overview}
    \vspace{-15pt}
\end{figure*}

\subsection{Challenges in Ultrasonic Delivery Modeling}\label{sec:ultrasound_observation}
The above attempts' failure drives us to look into the root cause of why modeling ultrasonic transformation is challenging. Ultrasound is intrinsically distinct from audible sounds due to its much higher frequency, and ultrasonic delivery leverages microphones' nonlinearity.
We also summarize the following characteristics:\\ 
\begin{itemize}[leftmargin=*,topsep=-20pt]
    \item \emph{Ultrasound-induced Noise:} 
    The ultrasound carrier continuously forces the diaphragm to vibrate, probably resulting in anomalous noise in recorded attack alike Fig.~\ref{fig:est_rir}\&\ref{fig:AE_sensitivity}. Combined with such variable ultrasound fields, a slight displacement (e.g., 3~cm) can lead to different audio patterns.
    \item \emph{Nonlinear Distortion:} Eq.~\ref{equ:nonlinearity} indicates the I/O relationship of nonlinear demodulation, in which the factors $k_i$ is unknown and varies with recording devices~\cite{li2023learning}. 
    \item \emph{Varying Soundfield:} As shown in Fig.~\ref{fig:soundfield}, ultrasound field (25~kHz) is significantly more directional and changes more dramatically than audible waves (1~kHz) due to the much shorter wavelength.
    \item \emph{Hardware-induced Instability:} Ultrasound-based attacks rely on a series of signal processing and sophisticated devices, thus bringing instability due to hardware imperfection. 
\end{itemize}

\section{Design of \alias}

\subsection{Overview}
\textbf{Design Goal.}
To manipulate ASRs while being used by users, adversaries shall create universal IAPs. However, they face the following challenges to obtain and deliver such perturbations:

\underline{\textit{Ultrasound Complexity (C1).}} Modeling the ultrasonic delivery is unprecedented compared to the audible-band RIR mimics, because ultrasound fundamentally differs from audible sound as listed in \textsection\ref{sec:ultrasound_observation}, including (i) ultrasound-induced anomalous noises, (ii) nonlinear distortion, (iii) varying sound field, and (iv) hardware-induced instability.

\blue{
\underline{\textit{User-ASR Connection (C2).}} ASR systems always respond to the user after receiving a command. Adversaries need to suppress the impact of \textit{user disruption}, i.e., break down the user-ASR connection by IAPs that can silence user's excessively long speech and remedy commands.}

\blue{
\underline{\textit{User Variation (C3).}} Since adversaries cannot exactly know the user speech's content, timing, or length, naively mixed speech signals will lead to undesirable ASR transcriptions. The tailored IAP needs to be universal while facing arbitrary user commands and superimposed time points.}

\underline{\textit{Physical Robustness (C4).}} The adversary also faces several factors that are variable in physical attacks, such as user loudness, hardware instability, and the user's environment (i.e., with different reverberations). We also extend the modulation method for reducing unexpected sound leakage.

\blue{To achieve adversaries' goal while addressing the aforementioned challenges, we propose \alias with unique technical design. This design includes: (1) tackling ultrasound complexity to deliver physically effective IAPs and therefore addressing \textit{user auditory} (cf.~\textsection\ref{sec:design_transformation}); (2) overcoming \textit{user disruption} to achieve real-time manipulation of ASR (cf.~\textsection\ref{sec:design_mute}, \ref{sec:design_universal}); (3) boosting attack stealthiness and practicality (cf.~\textsection\ref{sec:design_robust}). The optimization workflow of \alias is exhibited in Fig.~\ref{fig:design_overview}.}

\textbf{Problem Formalization.}
Unlike the audible-band AE attacks subject to stealthiness constraints, we achieve inaudible perturbations delivery using ultrasound modulation. Thus, we avoid the narrow constraints in Eq.~\ref{equ:audible_ae_formula}, e.g. $\epsilon<0.01$, where the IAP's optimization space can reach the maximum upper bound: $\delta\in[-1,1]^n$. We believe a broad optimization space possesses more feasible solutions, facilitating a universal attack. Combined with our core objective: fooling ASRs to recognize the superimposed speech of user voice and perturbation $x+\delta$ as the adversary-desired transcription $y_t$.
This basic idea can be optimized via the following formulation:
\begin{equation}
\begin{aligned}
    & {minimize}~ \mathcal{L}(f(x+\delta), y_t)\\
    & {s.t.}~ \delta \in [-1,1]^n ~{and}~ x+\delta \in [-1,1]^n
\end{aligned}
\label{equ:inaudible_formula}
\end{equation}



\subsection{Ultrasonic Transformation Modeling}\label{sec:design_transformation}
As shown in Fig.~\ref{fig:design_ufr}, our modeling exploits the \circled{5}additive property of the baseband audio $m$'s nonlinear transformation $H(f)m(f)$ and the ultrasound-induced anomalous noise $n$, and then \circled{6}yields estimated audio $\hat{m}=H(f)m(f)+n$ that is highly similar to the actual recorded audio $\widetilde{m}$.
In this subsection, we elaborate on our divide-and-conquer strategy of implementing ultrasonic transformation modeling that overcomes problems (i)-(iii) corresponding to \textit{ultrasound complexity (C1)}. Based on this, we can deliver physically effective IAPs via the steps \circled{1}$\sim$\circled{3} in Fig.~\ref{fig:design_overview}. We address the problem (iv) in \textsection\ref{sec:design_robust}. 

\subsubsection{Tackling Anomalous Noises}\label{sec:design_noise}
Ultrasound-based attacks modulate the baseband $m$ with $s(t)={A}{[1+m(t)]}c(t)$, where regardless of the energy of $m$, the carrier signal $c(t)=cos\omega_{c}t$ always emits and forces the microphone diaphragm into vibration, appearing abnormal noises~\cite{li2023learning}. 
Our experiment also demonstrates that although the recorded $s$ varies with $m$, the anomalous noise pattern is almost decided by the carrier. 
The nature of the ultrasound field further results in noise variation with different injection angles $\theta$ and distances $d$, showing irregular patterns. Therefore, due to such variation, neural networks fail to learn a stable mapping of the digital-to-physical domain.
We denote the noise $n(\theta,d)=f_n(\theta,d,s)$, where $f_n$ is the projection of the ultrasound signal $s$ to the recorded abnormal noise $n$ at different positions. 
In practice, an attacker can sample the variable anomalous noises by simply emitting the ultrasonic carrier. We collect a lightweight noise dataset using 25~kHz ultrasound (without modulation) of 1m at varying angles, forming a set $U_n$ of 25 pieces of 10-second noises.

\subsubsection{Ultrasonic Frequency Response}\label{sec:design_ufr}
\begin{figure}[t]
	\centering
	\includegraphics[width=0.48\textwidth]{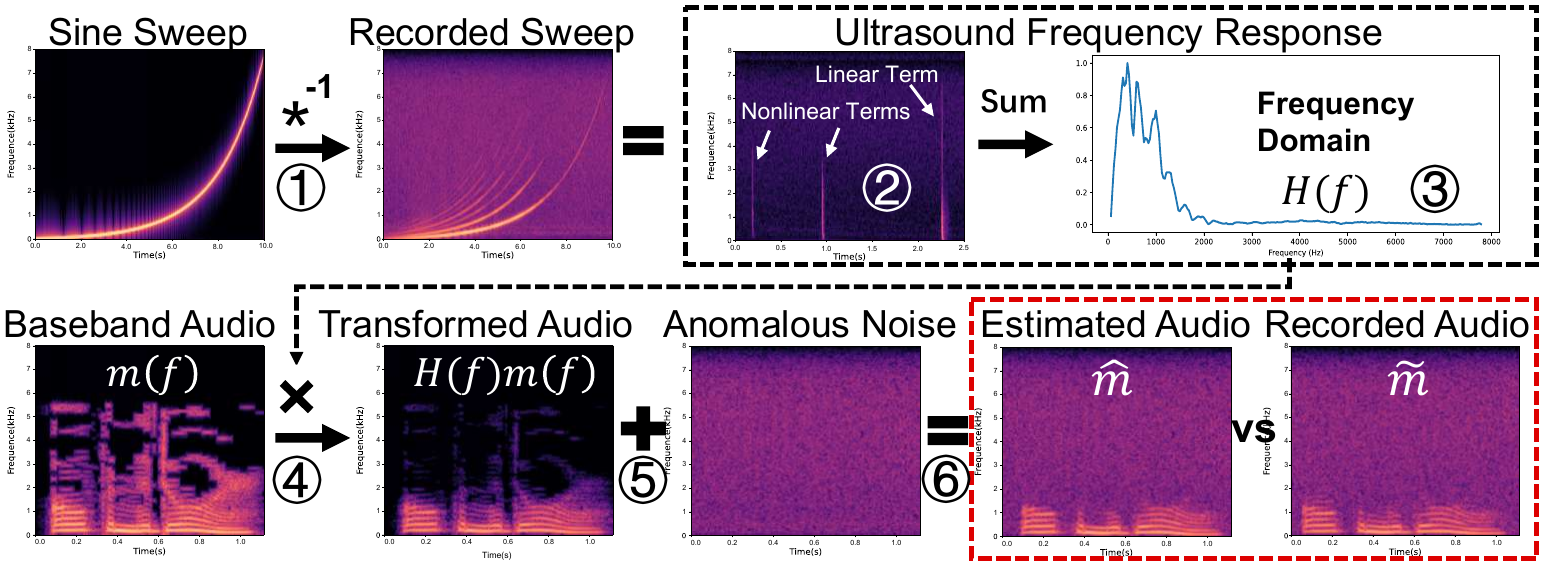}
	\caption{\label{fig:design_ufr}Ultrasonic transformation modeling. \textbf{1st row:} Procedure to obtain the ultrasonic frequency response (UFR). \textbf{2nd row:} High similarity between the estimated audio and actually recorded audio (the red box) proves its effectiveness.}
	\vspace{-15pt}
\end{figure}

Recalling the reasons for LTI system-based RIR's failure in \textsection\ref{sec:classic_rir} of modeling unprecedented ultrasonic delivery, except for anomalous noises, the inability to describe the nonlinear demodulation process is also a key factor. 
The adversaries aim to achieve robust and adaptive attacks with minimal effort, i.e., building an efficient transformation that can well estimate the demodulated pattern of a given digital perturbation after ultrasonic delivery in Fig.~\ref{fig:design_ufr} (red box). 
Fig.~\ref{fig:design_ufr} also depicts the recorded audio derived after inaudible signal injection, whose energy is clearly concentrated in the low-frequency band compared to the original audio~\cite{roy2018inaudible}.
Although nonlinearity exists, we are driven to obtain an ultrasonic frequency response (UFR) that characterizes the inaudible acoustic energy conversion at different frequencies.

We first overcome ultrasound-induced noises that hinder us from obtaining an accurate frequency response by adopting the sine sweep technique~\cite{farina2007advancements}, which can ignore components uncorrelated to the sweep signal during processing. We use it to generate a fast 10s sweep ranging from 50$\sim$7800~Hz, which is carefully chosen for diminishing hardware imperfection, and record it on the receivers, shown in Fig.~\ref{fig:design_ufr}\circled{1}. 
Thus, we can obtain the UFR $H(f)$ by deconvolution ($\ast^{-1}$). Notably, as shown in Fig.~\ref{fig:design_ufr}\circled{2}, it does shield the effects of noises and focuses on the frequency response measurement, which decouples the linear and nonlinear terms. We sum these terms up in Fig.~\ref{fig:design_ufr}\circled{3}, forming a holistic frequency-domain UFR of the received perturbations $\delta$ as $\overline{\Delta}(f)=H(f)\Delta(f)$, where $\Delta(f)=\mathcal{F}(\delta(t))$; $\mathcal{F}$ means Fourier Transform.


\subsubsection{Enabling Location-Variable} 
Uneven ultrasound field makes MLP-based method in \textsection\ref{sec:classic_nn} difficult to estimate transformation from arbitrary-position attack. As for efficient UFR, we believe that combining it with ultrasound $s(d,t)$ propagation process~\cite{holm2019waves} will empower to render more adaptive attacks:
\begin{equation}\label{equ:dis_filter}
    H(f,d) = H(f)\cdot e^{-a_0{\omega_c}^n d},~n\in[1,2]
\end{equation}
where $a_0$ is a medium-dependent attenuation parameter, $\omega_c$ is the carrier's frequency.
\blue{
Moreover, the energy variation caused by different injection angles is hard to model under such a changing sound field. We overcome this issue by conducting sine sweeps at different angles $\theta$ similar to \textsection\ref{sec:design_noise} and get 25 pieces of 10-second sweep clips.
Consequently, the collection of a complete set of UFRs and anomalous noises for subsequent optimization requires approximately 8.3 minutes.
Overall, with a pair of UFR $H_{\theta}(f,d)$ and noise clip $n(\theta,d)$ from the same location, we can well estimate the digital perturbation into its recording. However, to obtain a location-variable perturbation, we shall modify the expression of Eq.~\ref{equ:inaudible_formula} and find the perturbation via robust training:
}
\begin{equation}
    \underset{\delta}{argmin} \underset{h_{\theta} \sim U_H, n \sim U_n}{\mathbb{E}}\left[\mathcal{L}(f(x+h_{\theta}(d)\ast\delta+n), y_t\right]
\end{equation}
where we use time-domain expression $h_{\theta}(d)\ast\delta$ to indicate the transformed perturbation's waveform, as $H_{\theta}(f,d)\Delta(f)=\mathcal{F}[h_{\theta}(t,d)\ast\delta(t)]$ obeys the time convolution theorem.
We randomly select the UFR $H_{\theta}(f,d)$ and noise $n$ pairs from $U_H$ and $U_n$ during the optimization process to mimic actually delivering the inaudible adversarial perturbation at different locations. 
As we fully take ultrasound's inaudibility advantages, the experiment results also validate the optimization space is large enough to craft a robust perturbation effective under varying UFRs and noises.


\begin{figure*}[t]
	\centering
	\includegraphics[width=0.8\textwidth]{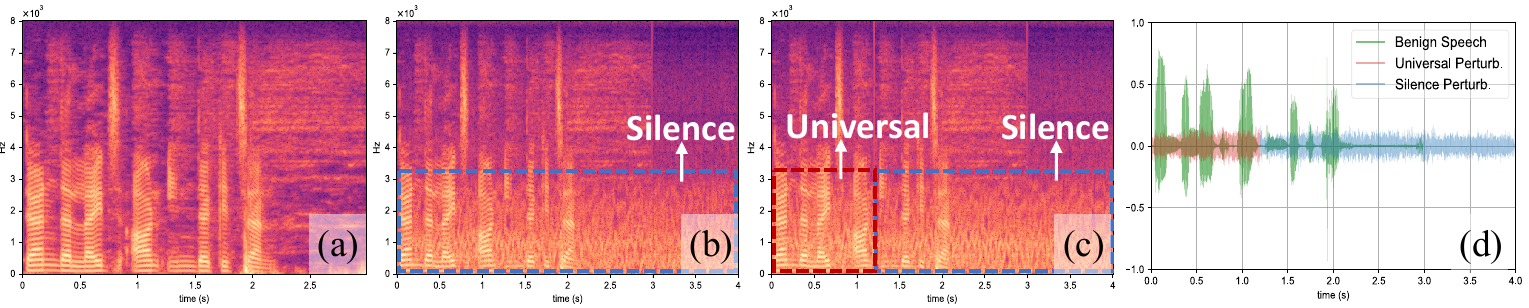}
        \vspace{-7pt}
	\caption{Diagram of \alias attacking a benign speech (a) with the silence (b) and universal goals in an \textit{alter-and-mute} manner (c\&d).}    
	\label{fig:design_attack}    
	\vspace{-15pt}
\end{figure*}

\subsection{\blue{Silence Perturbation}}\label{sec:design_mute}
\blue{Given the failures of prior works faced with \textit{user disruption}, specifically \ding{203} the challenge of excessively long instructions, and \ding{204} the potential counteraction through remedy commands, we believe that the solution lies in silencing the user instructions, i.e., breaking down the user-ASR connection when necessary.
Based on our ultrasonic transformation modeling, adversaries can materialize physically effective silence perturbations.
These perturbations can alter arbitrary user instructions to blank (`` '') in a targeted manner, effectively rendering the ASR system unresponsive to the user instructions.
We observe that implementing silence perturbations offers several advantages, including:
(1) When altering long user commands with short target intent, such as ``start recording'' (case \ding{203}). The silence perturbation can be linked alongside the universal perturbation in an \textit{alter-and-mute} fashion (cf. \textsection\ref{sec:design_universal}) so that the ASR will output adversary-desired transcription;
(2) it guarantees that users cannot meddle in running malicious operations by issuing remedy commands (case \ding{204}), even if they notice the presence of attacks; (3) it can render ASR services in the denial-of-service condition, preventing users from using them normally.}

Fig.~\ref{fig:design_attack}(b) depicts the diagram of a robust silence perturbation $\xi$, which is expected to superimpose over any benign content like Fig.~\ref{fig:design_attack}(a) and leads the final transcription of ASR to blank $y_b$ (`` '').  The length of $\xi$ is empirically set to 5s based on our experiments, for which we balance the duration of common speech instructions and the optimization overhead. For the case of excessively long user utterances, we address them in the generation process by repeating the perturbation.
\blue{To craft such a content-agnostic $\xi$, we improve the penalty-based expectation function to find the silence perturbation over a group of common voice commands $U_x$, as shown in Fig.~\ref{fig:design_overview}\circled{4}.}
\begin{equation}
    \underset{\xi}{argmin} \underset{h_{\theta} \sim U_H, n \sim U_n, x\sim U_x}{\mathbb{E}}[\mathcal{L}(f(\mathcal{S}_{x}+{h_{\theta}(d)\ast\xi+n}), y_b)]
\end{equation}
\blue{where $\mathcal{S}_{(\cdot)}$ means randomly shifting the user utterances $x$ for introducing randomness to the superimposed time within a preset $T$:100~ms. $U_x$ is elaborated in the experimental setup \textsection\ref{sec:eval_dataset}. It is more practical than the case where an AE and user speech are required to be perfectly aligned. The details of content-agnostic and synchronization are given in \textsection\ref{sec:design_universal}.}
\subsection{Universal Perturbation}\label{sec:design_universal}
\blue{Different from the proof of concept of universal AEs against a CNN-based speech command classification model presented in~\cite{li2020advpulse} by exploiting the temporal insensitivity of CNNs, the RNN-based models widely deployed on commercial ASR services are more difficult to attack.
This difficulty arises because end-to-end ASRs, such as DeepSpeech~\cite{amodei2016deep}, employ connectionist temporal classification (CTC) that calculates the loss between a continuous speech feature sequence and a target transcription, making it context-dependent. Consequently, when introducing subtle perturbations in different contexts, it is often difficult to ensure that the CTC losses of multiple mixed signals will simultaneously converge to the desired target.}

\subsubsection{\blue{Content-Agnostic}}
\blue{We believe that the reasons why previous audible-band AEs struggle to tamper with large amount of speech content are two-fold: \textit{user auditory} and \textit{user disruption}. 
To avoid being noticed by users, prior adversarial perturbations are limited by imperceptibility constraints and signal forms (e.g., with short length and subtle amplitude). Consequently, these perturbations are fragile and easily defensible. In contrast, our IAP delivery is completely inaudible via ultrasound modulation. Thus, the perturbation's length and amplitude are unconstrained, maximizing its optimization space. We fully use the advantages to generate a universal perturbation that can alter substantial short utterances into adversary-desired intent, e.g., a 1.2s $\delta$ tailored for “open the door”.}

\blue{
However, for excessively long speech or possible subsequent remedy commands in user-present scenarios (\textit{user disruption} \ding{203}-\ding{204}), the adversary should resort to the silence perturbation in \textsection\ref{sec:design_mute}, which can cooperate well with the universal perturbation in an \textit{alter-and-mute} manner.
As depicted in Fig.~\ref{fig:design_attack}(c) and (d), when the universal perturbation $\delta$ is combined with a well-trained silence perturbation $\hat{\xi}$, the former can apply to alter the user commands, and the latter will mute the subsequent user commands or remedies.}
As illustrated in Fig~\ref{fig:design_overview}\circled{4}, we determine the optimal $\delta$ by optimizing the following expectation function:
\begin{equation}
    \underset{\delta}{argmin} \underset{h_{\theta} \sim U_H, n \sim U_n, x\sim U_x}{\mathbb{E}}[\mathcal{L}(f(x+{h_{\theta}(d)\ast\overline{\delta:\hat{\xi}}+n}), y_t)]
    \vspace{-2pt}
\end{equation}
\blue{where {\scriptsize$\overline{\delta:\hat{\xi}}$} means the universal perturbation $\delta$ followed by a crafted silence perturbation $\hat{\xi}$. $U_x$ is the same subset used for generating silence perturbations, whose details are given in  \textsection\ref{sec:eval_dataset}.}


\subsubsection{Synchronization-Aided}\label{sec:sync_free}
Although the universal perturbation can deceive the ASR with any victim’s speech into adversary-desired intent, an adversary can hardly deliver attacks synchronously when the victim vocalizes. 
\blue{Out of attack practicality, we propose a VAD-based synchronization mechanism to achieve real-time manipulation, which avoids continuous AE broadcasting or assuming an adversary always ready for attacking.
Specifically, we employ a microphone to record the user's voice. Once detecting the user's speech via voice activity detection (VAD), our program automatically triggers the emission of the prepared perturbation. 
Based on our experiments, the delay is impacted by three stages of our real-time pipeline: (1) from user vocalizing to being detected by the running VAD program (5$\sim$20~ms); (2) software-to-hardware IAP triggering (5$\sim$15~ms); (3) ultrasound propagation (0$\sim$30~ms). Due to the delay uncertainty, we consider bringing the time randomness of a range $T$ into our optimization, whose upper bound is empirically set to 100~ms. For a direct reference, the average overall delay when attacking at 4m is around 27~ms, far below the maximum tolerable delay (100~ms) preset during optimization.
Particularly, the recording of user speech can also be utilized to present a more covert attack by inaudibly replaying user-desired commands, as ``\textit{Man-in-the-middle Attack}'' stated in \textsection\ref{sec:threat_model}.
By integrating the above-mentioned optimization objectives, we further craft the universal IAP through the below expectation:}
\begin{equation}\label{equ:sync_attack}
    \underset{\delta}{argmin} \underset{h_{\theta} \sim U_H, n \sim U_n, x\sim U_x}{\mathbb{E}}[\mathcal{L}(f(x+\mathcal{S}_{h_{\theta}(d)\ast\overline{\delta:\hat{\xi}}+n}), y_t)]
\end{equation}
where $\mathcal{S}_{(\cdot)}$ mimics \alias can be superimposed on victim speech at random time points (Fig.~\ref{fig:design_overview}\circled{4}) within the preset $T$.

\subsection{Physical Robustness}\label{sec:design_robust} 
\subsubsection{Loudness Adaptive and Hardware Instability} When conducting physical attacks, \alias is able to handle the challenges of ultrasound nature based on our digital-to-physical transformation in \textsection\ref{sec:design_transformation}. However, the loudness of the victim's speech varies with context or emotion, and hardware instability still exists.
These will result in difficulty maintaining our inaudible perturbation in effectively altering the victim's voice if the mutual energy relationship between the two is inconsistent with the optimization process.
As shown in Fig.~\ref{fig:design_overview}\circled{5}, we introduce relative volume augmentation into the crafting process, which exploits a hyper-parameter $\beta$ denoting a range of user speech's volume and thereby brings randomness to the mutual relationship between user voice and perturbation.

\subsubsection{Attack at Different Environments} Although ultrasound-based attacks directly inject into recording devices' microphones and are reverberation-free, the audible-band human voice still goes through multi-path reflections and ambient noises in different environments. To alter user commands regardless of the impact of scenes, we apply random RIR and noise clips from the Aachen Impulse Response (AIR) Database~\cite{jeub2009binaural} in Fig.~\ref{fig:design_overview}\circled{6}, including small, medium, large rooms and corridors for user speech augmentation.

\subsubsection{\blue{Single-Sideband Extension}}\label{sec:SSB}
\blue{
Although \alias can achieve real-time manipulation of the ASR output very covertly using sophisticated devices (e.g., narrowband ultrasonic transducers and signal generators) at long distances through windows or doors, we aim to accomplish highly stealthy attacks even in close proximity to the victim by utilizing everyday-life loudspeakers or portable attack devices.
However, the simple amplifiers, sound cards, and off-the-shelf loudspeakers exhibit poor suppression of intermodulation and harmonics of high-frequency DSB-AM signals. Namely, they present increased nonlinearity, resulting in sound leakage (cf. \textsection\ref{sec:discuss}). 
To enable attacks with portable devices and loudspeakers (cf. \textsection\ref{sec:portable_attack}), we adopt single-sideband amplitude modulation (SSB-AM), which removes one of the sidebands based on the Hilbert transform~\cite{ziemer2006principles}. 
Compared to DSB-AM, SSB-AM has only half the bandwidth, rendering higher transmission efficiency. 
Importantly, it mitigates the intermodulation between different sideband frequencies, making the sound less prone to leakage than DSB-AM at the same energy level. Specifically, we employ upper sideband modulation (USB-AM), formalized as $S(t)=m{cos\omega_{c}t}-\hat{m}{sin\omega_{c}t}+cos\omega_{c}t$, rather than lower sideband modulation (LSB-AM), as the former exhibits better inaudibility in our experiments and more details are given in Appendix  \textsection\ref{append:SSB}.}

Overall, the algorithm of \alias is described in Algorithm \ref{algo1}, Appendix \textsection\ref{append:algo1}, where we demonstrate the optimization process of crafting \alias from scratch.

\section{Evaluation}
\subsection{Experiment Setup}
\subsubsection{Overview} We implement \alias using PyTorch~\cite{pytorch} on a Ubuntu 20.04 server with Intel Xeon 6226R 2.90GHz CPU and NVIDIA 3090 GPU. Based on our experiments, we empirically set the default configuration as $\delta=1.2s$, $\xi=5s$, $\epsilon=1$, $0.5\le\beta\le1.5$, sync range $T$=100~ms, $maxEpoch$=800. Adam optimizer~\cite{kingma2014adam} is used to speed up our convergence.
For evaluating \alias's effectiveness in fooling ASR while users use it, we select the end-to-end DeepSpeech2~\cite{amodei2016deep} as the target model and conduct experiments in both digital and physical scenarios.

\subsubsection{Dataset}\label{sec:eval_dataset} We adopt the typical Fluent Speech Command Dataset~\cite{fluent2020commands} to examine the effectiveness of \alias, including 30,046 voice command samples. 
We randomly selected 896 samples from the 10-person validation set given in the dataset, with each speaker contributing around 90 utterances on average. These samples are used to craft our perturbation. The remaining unseen 29,150 samples are used to evaluate \alias under various settings. 


\subsubsection{Hardware} We employ a signal generator (SIGLENT SDG6032X)~\cite{siglent} to modulate the created IAPs, a power amplifier (NF HSA4015)~\cite{amplifierHSA4015} to enable long-range delivery, and a custom ultrasound transducer array to emit the modulated IAPs. The recording devices to be tested include Google Pixel 3aXL, iPhone14 pro, MI Mix2s, OPPO Reno5 pro, and ReSpeaker Mic array v2.0~\cite{respeaker}, where all model versions are released in the last five years. Moreover, we evaluate attacks with a self-made portable device and a loudspeaker in \textsection\ref{sec:portable_attack}.

\subsubsection{Metrics} (1) We use the success rate (SR) to indicate the percentage of \alias successfully altering user commands and matching target transcriptions in all attempts. (2) We use character error rate (CER), a representative metric in ASR tasks, to indicate the adversary's ability to tamper with user commands from the character level; a lower CER represents a more effective attack. (3) Signal-to-Noise Ratio (SNR) and $L2$-distortion are vital for audible-band AEs because of the imperceptibility requirements. SNR: the ratio of benign audio power to the perturbation power. $L2$: the sum of squared amplitude. AEs with a low SNR and high $L2$ are more likely to be noticed, and vice versa.



\subsection{Digital Attack Performance}
As our attack focuses on real-world scenarios, where physical disturbances always exist, we incorporate the effects of physical conditions by employing our ultrasonic transformation modeling to guarantee that digital attack performance has physical significance. 

\subsubsection{Impact of Optimization Space} 
\begin{figure}[!tb]
	\centering  
        \subfigure[Silence Perturbation]{   
    	\begin{minipage}[t]{0.225\textwidth}
    		\centering
    		\includegraphics[width=1\textwidth]{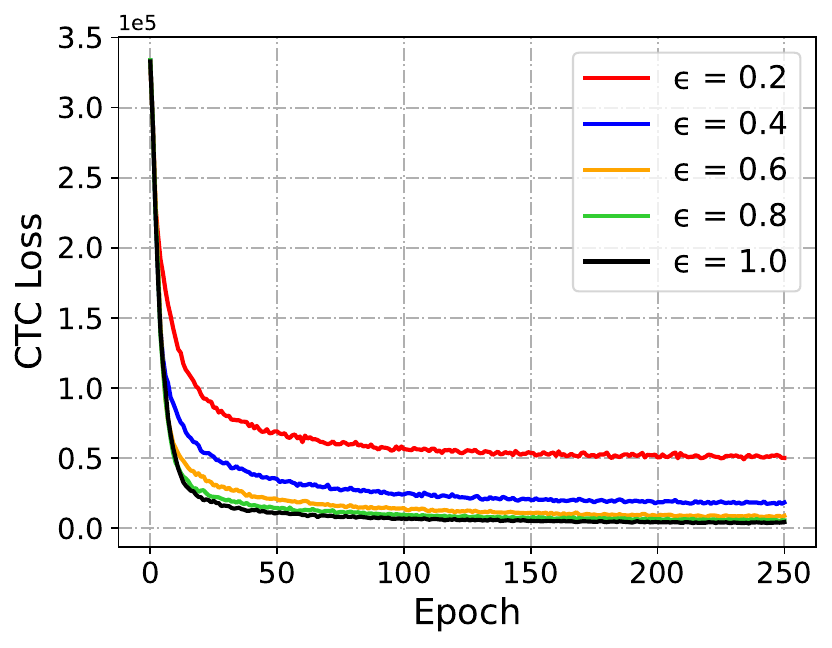} 
    	\end{minipage}
        }
	\subfigure[Universal Perturbation]{   
	\begin{minipage}[t]{0.225\textwidth}
		\centering
            \hspace{-0.1\linewidth}
		\includegraphics[width=1\textwidth]{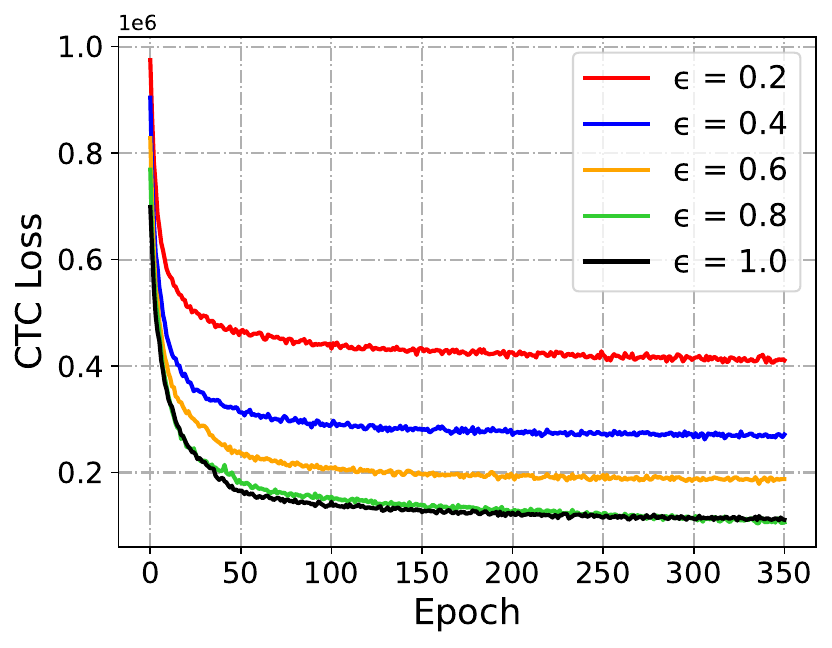} 
	\end{minipage}
	}
	\caption{CTC loss curves of silence and universal perturbations during the optimization process under varying $\epsilon$.}    
	\label{fig:dig_exp1}    
	\vspace{-10pt}
\end{figure}

Since the delivery of \alias is inaudible, it facilitates the unconstrained advantage of setting $\epsilon$ up to $1$ (i.e., the normalized audio's upper bound) for universal attacks. We further explore attack capability under different $\epsilon$ upper bounds, both universal and silence IAPs.
We optimize silence perturbations according to $\epsilon=0.2, 0.4, 0.6, 0.8, 1.0$, respectively, aiming to tamper the user instructions to the blank. In addition, we obtain the universal perturbations expected to alter user commands to ``open the door'' with the same settings. CTC loss convergence curves are shown in Fig.~\ref{fig:dig_exp1}. We observe that the crafting process can converge faster as the $\epsilon$ (i.e., the optimization space) increases in both tasks. After $\epsilon$ reaches $0.8$, the convergence rate approaches the maximum.
Then we estimate the physical delivery of both perturbations via an unseen transformation model (i.e., a pair of UFR and anomalous noise not involved in training). The transformed perturbations are further superimposed on every test voice command sample. Results listed in Tab.~\ref{tab:dig_exp1} show that, in addition to the faster convergence, a larger $\epsilon$ significantly boosts the universality of \alias. It can successfully alter 18,946 samples into ``open the door'' and mute 27,531 user commands into blank `` '', which highlights \alias features a highly universal capability.  

\begin{table}[t]\footnotesize
	\centering
		\caption{The number of successfully silenced/altered test speech samples under different $\epsilon$ upper bounds}
		\renewcommand\arraystretch{0.8}
		\renewcommand\tabcolsep{5pt}
		\begin{threeparttable}
			\begin{tabular}{c|c|c|c|c|c}
				\toprule
				\textbf{Upper Bound ($\epsilon$) } & \textbf{$0.2$} & \textbf{$0.4$} & \textbf{$0.6$} & \textbf{$0.8$} & \textbf{$1.0$} \\
				\midrule
				\textbf{Silence Perturb.} & 1,591 & 8,095 & 17,064 & 24,832 & 27,531 \\ \midrule
				\textbf{Universal Perturb.} & 649 & 5,268 & 13,085 & 16,726 & 18,946 \\
				\bottomrule
			\end{tabular}
		\end{threeparttable}
		\label{tab:dig_exp1}
\end{table}

\subsubsection{Comparison of Convergence Overhead and Audibility Cost for the Universality Goal}\label{eval:dig_compare} 
\begin{figure}[t]
  \centering
  \begin{minipage}[b]{0.235\textwidth}
    \centering
    \hspace{-0.1\linewidth}
    \includegraphics[width=\textwidth]{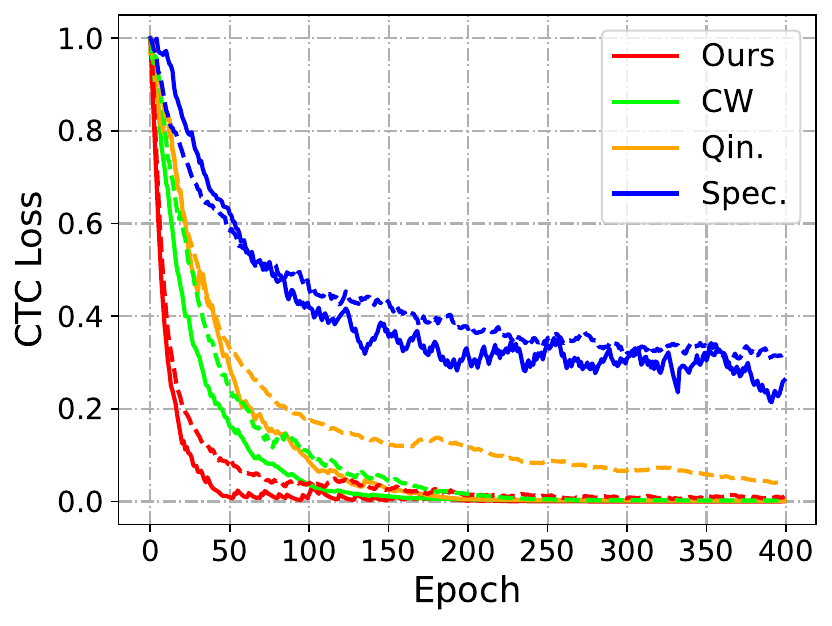}
  \end{minipage}
  \hfill
  \begin{minipage}[b]{0.235\textwidth}
    \centering
    \footnotesize
    \renewcommand\arraystretch{0.7}
    \renewcommand\tabcolsep{2.pt}
    \hspace{-0.1\linewidth}
    \captionof{table}{Audibility cost}\vspace{-10pt}
    \raisebox{0.35\linewidth}{
        \begin{tabular}{c|c|c|c}
            \toprule
            \textbf{Method~} & \textbf{~Goal~} & \textbf{~~SNR~~}  & \boldmath{~~$L2$~~} \\ \midrule
            \multirow{2}{*}{CW}    & 1    & 22dB & 1.31 \\ & 5    & 18dB & 2.24 \\ \midrule 
            \multirow{2}{*}{Qin}   & 1    & 22dB   & 1.42 \\ & 5  & 17dB   & 2.67 \\ \midrule 
            \multirow{2}{*}{Spec.} & 1    & 8.5dB  & 122 \\ & 5   & 8.4dB   & 123 \\ \bottomrule 
        \end{tabular}
        \label{tab:dig_exp2}
    }
  \end{minipage}
  \normalsize
  \caption{CTC loss curves. Compare the convergence speed of \alias with 3 classical audible-band AEs. Dashed lines: train a perturbation that can simultaneously alter 5 voice commands; likewise, Solid lines: alter 1 command, thus converging faster than the former.}
  \label{fig:dig_exp2}
  \vspace{-12pt}  
\end{figure}

\blue{The unconstrained advantage of \alias ($\epsilon=1$) empowers its high universality. We further compare it with 3 classical audible-band AEs (i.e., CW~\cite{carlini2018audio}, Qin~\cite{qin2019imperceptible}, and SpecPatch~\cite{guo2022specpatch}) regarding the cost for achieving the same universality goal of each method. We reproduce these works strictly following their instructions.
We set \textit{two goals}: creating a single perturbation that can alter 1) one or 2) five commands based on each method. Notably, for audible-band AEs, we employ RIRs for physical simulation to be consistent with our default setup.
We specify the minimal upper bounds $\epsilon$ of CW, Qin, and SpecPatch to $0.03,0.05,0.05$, respectively, based on which the three methods can maintain universality for 5 commands, i.e., finally converge to the target transcript ``Open the door''. 
We examine also the CTC loss convergence speed of 4 methods. The normalized loss curves in Fig.~\ref{fig:dig_exp2} clearly show that \alias (in red) converges within the fewest iterations among 4 methods; SpecPatch (in blue) converges slowest as it is devised to be short (0.5s). Specifically, we list the overall duration for each methods to final convergence for altering 5 commands---\alias: 1.63~min, CW: 6.52~min, Qin: 9.16~min, and SpecPatch: 35.38~min. \alias converges faster because (1) we reduce optimization complexity by only picking 5 random UFR/noise pairs rather than all ultrasonic channel data per iteration; (2) \alias can quickly find feasible solutions due to its broad optimization space.}
In addition, Tab.~\ref{tab:dig_exp2} demonstrates the SNRs and $L2$-distortion values of audible-band AEs under different universality goals. All SNRs of these AEs are low due to a compromise of physical robustness and imperceptibility, with the highest SNR down to 22 dB. Moreover, if the goal number increases, audible-band AEs are bound to get louder and more easily heard.

\subsubsection{Different Target Commands}\label{sec:eval_commands}
Given that adversaries may launch attacks for various purposes, they will craft different adversarial perturbations accordingly.
In this experiment, we first train 10 universal perturbations referring to typical malicious commands~\cite{petracca2015audroid} listed in Appendix \textsection\ref{append:command_list} Tab.~\ref{tab:diff_commands} along with the silence perturbation. Then we apply \alias to 7,200 benign commands to validate its effectiveness, amounting to 72,000 samples. 
We count the success rate when transcription outputs match the target commands correctly. In addition, we also count CERs over all samples. We find no significant performance varying with target transcripts, where most targets derive a 100\% SR and 0\% CER (7 out of 10). The lowest SR is still up to 92.82\%, corresponding to ``Mute volume and turn off the WiFi''. Moreover, it is worth noting that the highest CER of these targets is still down to 0.50\%, suggesting \alias can tamper with user commands well from the character level. Due to page limitations, the details are listed in Appendix~\ref{append:command_list} Tab.~\ref{tab:diff_commands}.


\subsection{Physical Attack Performance}
We perform extensive physical experiments to evaluate the practical performance of \alias under different conditions, i.e., w/o our modeling, distances, environments, recording devices, etc.
In the physical experiments, 
we set the target intent as ``open the door'', the attack distance 4m away from the recording devices with the injection angle pointing to their bottom microphones as the default configuration unless otherwise specified. Except for the experiments about different scenes, the rest are conducted in a laboratory of approximately 13.6m$\times$5.2m with slight HVAC noises. We employ a custom ultrasonic transmitter for inaudible adversarial perturbation delivery. A loudspeaker plays the audible benign speech samples, and the ambient noise level is around 38~dB. 
\blue{We also deploy a VAD-based program in conjunction with a microphone connected to the laptop to trigger IAPs delivery using the synchronization-aided design. This ensures real-time triggering when audible benign speech initiates.}
Our real-world attack scenario is given in Appendix \textsection\ref{append:attack_scenario}, Fig.~\ref{fig:attack_actual}. 

\subsubsection{Ablation Experiments w/o Transformation Modeling}\label{sec:eval_utm}
To validate the effectiveness of our ultrasonic transformation modeling, we apply 3 strategies to craft IAPs. In addition, we apply direct ultrasound-based attacks as the baseline group (\textit{G1}).
The first strategy is an optimization without transformation, i.e., $h_{\theta}(d)\ast\delta+n$ in Eq.~\ref{equ:sync_attack} is degraded to simple $\delta$ during the crafting process (\textit{G2}). 
Similarly, the second strategy uses a low-pass filter, reducing the precise transformation to a filter that allows signal components below 3~kHz to pass (\textit{G3}).
The third strategy is crafting a perturbation with \emph{our transformation}, i.e., \alias (\textit{G4}). 
We carry out experiments with synchronization-aided emission of both benign audio and attacks. We select 40 benign utterances to be played via loudspeaker, and finally collect 480 mixed samples (120 per group) by repeating the operation three times for minimizing errors. Tab.~\ref{tab:wo_transformation} lists each group's success rate and average CER, which remarkably denotes that our modeling can well describe the digital-to-physical transformation during optimization. Approximating the transformation as a low-pass filter can also generate a physically available perturbation with 21.67\% SR and 19.39\% CER. The attack performance decreases to 0\% in \textit{G1} and \textit{G2}. However, attacks without modeling still outperform the baseline (\textit{G1}) from the CER perspective due to leveraging the model vulnerability.

\begin{table}[t]\footnotesize
	\centering
		\caption{Ablation of w/o transformation modeling}
		\renewcommand\arraystretch{0.7}
		\renewcommand\tabcolsep{2pt}
		\begin{threeparttable}
			\begin{tabular}{c|c|c|c|c}
				\toprule
				\textbf{Metrics} & \textbf{Baseline (\textit{G1})} & \textbf{Without (\textit{G2})} & \textbf{Low-pass (\textit{G3})} & \textbf{With (\textit{G4})} \\
				\midrule
				\textbf{SR} & 0\% (0/120) & 0\% (0/120) & 21.67\% (26/120) & 100\% (120/120) \\ \midrule
				\textbf{CER} & 95.7\% & 78.93\% & 19.39\% & 0\% \\
				\bottomrule
			\end{tabular}
		\end{threeparttable}
		\label{tab:wo_transformation}
\end{table}

\begin{figure}[t]
    \centering
    \includegraphics[width=0.45\textwidth]{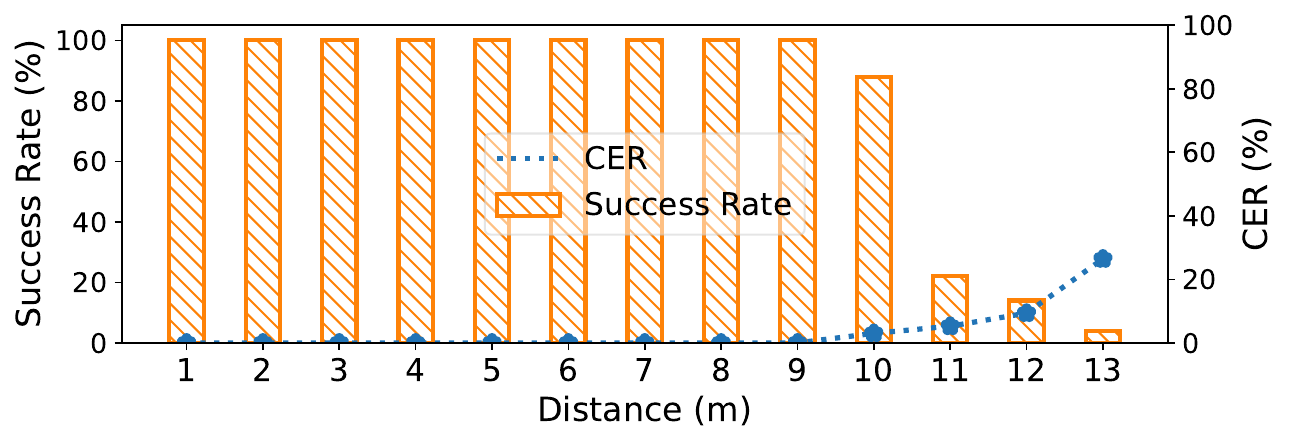}
    \caption{\alias's performance at different distances.}
    \label{fig:attack_distance}
    \vspace{-10pt}
\end{figure}
\subsubsection{Different Attack Distances}
Attacks in the audible band are constrained by concealment, resulting in perturbations that cannot be delivered with more extensive ranges. By contrast, our attack delivery via ultrasound modulation can apply substantial power, which overcomes the attenuation nature of ultrasound. We adjust the amplifier gain so that the high-frequency beam's energy reaching microphones is maintained, thus ensuring the effectiveness of \alias. Specifically, we conduct experiments at the ultrasonic transmitter away from the receiving device within 1m$\sim$13m (1m interval), where the maximum power at 10m$\sim$13m is approximately 3.2 Watt. We randomly select 40 voice commands and play them at each location. We repeat the perturbation superimposed on the benign command 3 times and totally collect 1,560 samples, with 120 per distance, respectively, as well as feed them into the ASR model. We count the success rate and CER in Fig.~\ref{fig:attack_distance}, where \alias is very effective within 1m$\sim$9m as the SRs are up to 100\% and CERs are down to 0\%. The SR is 88.7\% and CER is still down to 3.25\% at 10m. Besides, we observe the attack performance decrease at 11m$\sim$13m. We believe this is due to the ultrasound attenuation, which makes the perturbation less significant to the ASR model. We also discuss this issue in \textsection\ref{sec:discuss}.

\subsubsection{Different Attack Angles}\label{append:eval_angles}
\begin{figure}[t]
    \centering
    \includegraphics[width=0.45\textwidth]{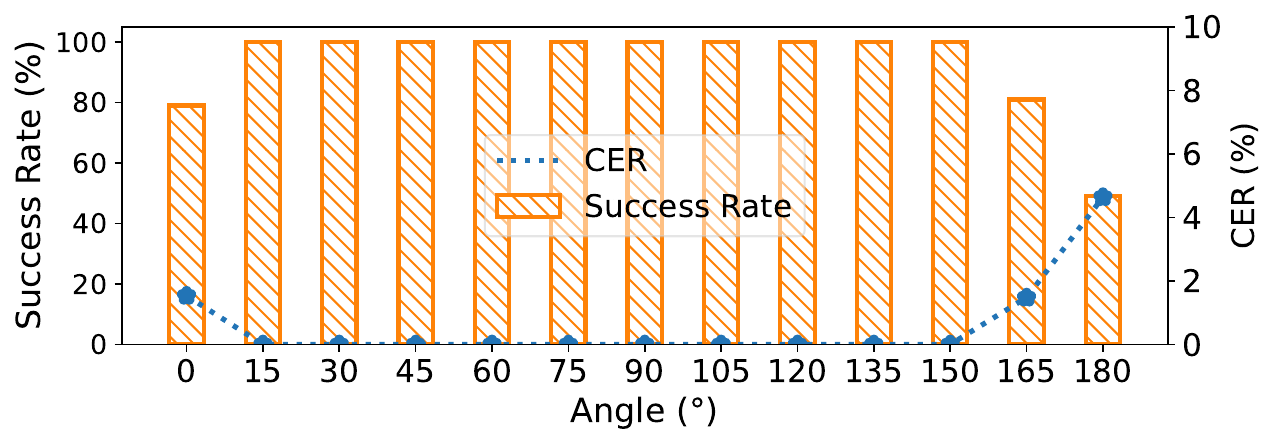}
    \caption{\alias's performance at different angles.}
    \label{fig:attack_angle}
    \vspace{-10pt}
\end{figure}

In this experiment, we keep the recording device's bottom microphone spatially within the ultrasound beam's coverage and set the attack distance to 2.5m. We rotate the recording device from 0\degree to 180\degree at 15\degree intervals, among which 90\degree means the ultrasound directly points to the bottom microphone. Under each angle, we play 40 benign commands and emit the universal IAP.
Eventually, we collected 520 mixed audio signals from 13 angles. As shown in Fig.~\ref{fig:attack_angle}, although ultrasound is highly directional, we find that there is no significant difference with 100\% success rate among different angles within 15\degree$\sim$150\degree. As the deployed location of bottom microphones varies with different phones, therefore attack performance is not symmetrical with angles (i.e., 79\% at 0\degree and 49\% at 180\degree). Overall, as most voice-interface devices nowadays are equipped with omnidirectional microphones, \alias can be effective as long as the beam can cover the bottom microphone. 

\begin{table}[t]\footnotesize
	\centering
		\caption{Different attack scenes}
		\renewcommand\arraystretch{0.7}
		\renewcommand\tabcolsep{3pt}
		\begin{threeparttable}
			\begin{tabular}{c|c|c|c|c}
				\toprule
				\textbf{Scene} & \textbf{Office} & \textbf{Lounge} & \textbf{Laboratory} & \textbf{Corridor} \\
				\midrule
				\textbf{SR} & 100\% (40/40) & 95\% (38/40) & 100\% (40/40) & 92.5\% (37/40) \\ \midrule
				\textbf{CER} & 0\% & 0.79\% & 0\% & 1.04\% \\
				\bottomrule
			\end{tabular}
		\end{threeparttable}
		\label{tab:diff_scene}
\end{table}

\subsubsection{Different Scenes}
To examine the effectiveness of \alias in different environments, our experiments include a small office (2.4m$\times$2.6m, 36~dB), medium lounge (6.3m$\times$3.8m, 42~dB), large laboratory (13m$\times$5.2m, 38~dB), and narrow corridor (60m$\times$2m, 44~dB). In these scenes, the reverberation pattern of audible sound varies with space size.
Our configuration consists of a transmitter-to-device distance of 4m and a loudspeaker-to-device distance of 1m, which mimics the standard user interaction distance, except the distance of 2.5m for the small office due to its limited size. 
We also play 40 audible benign samples and superimpose \alias on them for once. Then we collected 160 samples from 4 spaces. As shown in Tab.~\ref{tab:diff_scene}, we find no significant difference between these scenarios, as our design considers such physical variation.

\begin{figure}[t]
    \centering
    \includegraphics[width=0.42\textwidth]{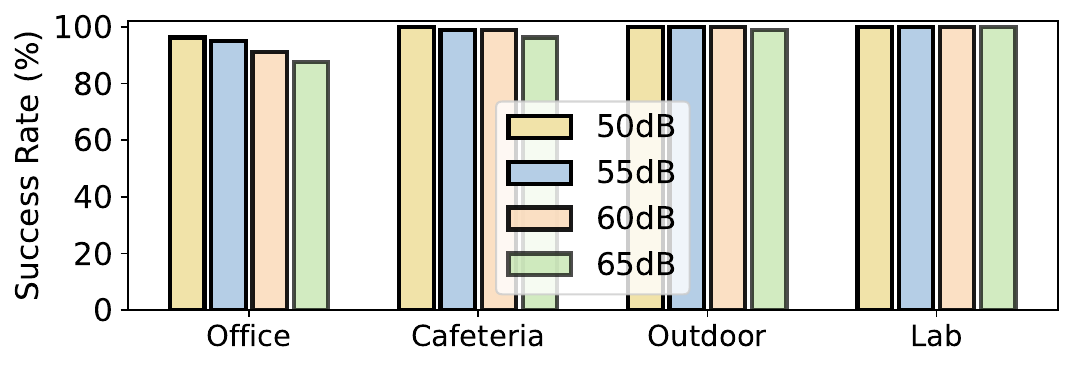}
    \caption{Attack performance in face of noises from typical scenes at 4 sound pressure levels.}
    \label{fig:eval_noises}
    \vspace{-10pt}
\end{figure}

\subsubsection{Different Ambient Noises}
We perform ambient noise-related experiments in our laboratory, where noises of 4 typical scenes are involved, i.e., cafeteria (people chatting), office (keyboard typing), lab (machine running), and outdoor (wind blowing) downloaded from the freesound~\cite{freesound}.
We evaluate noise starting from 50$\sim$65~dB, with 5~dB intervals, and we play noises through an additional loudspeaker to guarantee the noise pressure level reaches the receiver at 50, 55, 60, and 65 dB. Noise samples from 4 scenes are played continuously. At the same time, we play 20 audible benign commands and deliver \alias. Given that the noise is not constant, the superposition of different parts may have different effects. We repeat the above operation three times and collect 240 mixed samples for each noise level. Fig.~\ref{fig:eval_noises} demonstrates that \alias maintains effectiveness even if the noisy ambient sound reaches 65~dB with an average SR up to 97.65\%. The performance drops slightly in the office noise case of 87.5\%, where the keyboard typing and mouse striking are crisp noises with intense high-frequency energy. Since \alias mainly affects low-frequency acoustic features after transformation, high-frequency noise might reduce its attack performance on deceiving ASR models. 


\begin{figure}[t]
    \centering
    \includegraphics[width=0.42\textwidth]{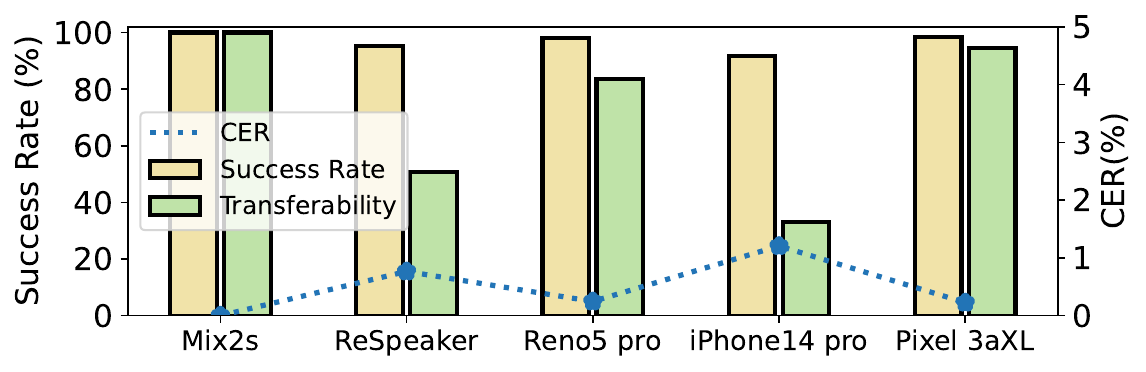}
    \caption{Attack performance on different recording devices.}
    \label{fig:phy_devices}
    \vspace{-10pt}
\end{figure}

\subsubsection{Different Recording Devices}
Since the ultrasound frequency response varies with different recording devices and microphone models~\cite{li2023learning}, i.e., we establish a specific transformation model for each device.
To verify that our perturbation can still manipulate the ASR model after being recorded by different devices, we obtain 5 pairs of universal and silence perturbations based on the device-wise ultrasonic transformation. 
After a similar collection as the above experiments done for each device, Fig.~\ref{fig:phy_devices} depicts the average SR of these devices is up to 96.8\% and CER is down to 0.50\%, where Mix2s reaches 100\% and 0\% on these metrics, proving the crafted \alias's effectiveness on individual devices. Moreover, \alias gets 95.8\% SR on ReSpeaker, suggesting it can also attack devices with multi-channel microphones well.
\blue{
Furthermore, given that adversaries may attack unmodeled (i.e., unseen) devices, we want to investigate \alias's transferability despite our ultrasound transformation modeling is device-specific. We apply the optimized perturbation of Mix2s to other devices. Among them, the Mix2s' combined perturbation can transfer to Pixel 3aXL and Reno5 pro with 94.2\% and 83.3\% SR. Besides, the performance reduces on iPhone14 pro (31.7\%) and ReSpeaker (50.8\%) due to their microphones' different frequency selectivity to ultrasound. The result indicates that \alias is also transferable across devices.}

\subsubsection{Different Speech \& Perturbation Loudness}
We further investigate the attack performance changes due to different loudness of the user speech and the universal perturbation. 
We set the representative audible sound pressure level to vary from 65$\sim$90~dB using a decibel meter and also vary ultrasonic emission power to keep the same loudness.
We play our perturbation, repeating 5 times at each volume level. Due to page limitation, results are given in Appendix \textsection\ref{append:loudness_compare}, Fig.~\ref{fig:loudness_compare}. As the mutual loudness changes, we find that once the perturbation has the same volume as the benign audio, it achieves over 55\% SR.
Moreover, with 5~dB higher than the benign audio, \alias can work effectively with an average SR up to 95.5\%. When \alias's volume is 10~dB higher than audible speech, it can dominate all the user commands. Notably, even if the direct ultrasound-based attack is 35~dB louder than the audible audio, the ASR model still recognizes a CER up to 46\%. In that case, \alias achieves all CERs down to 0\%.

\subsection{Attack with Portable Device and Off-the-shelf Loudspeaker}\label{sec:portable_attack}
Our sophisticated device facilities an extensive attack range, providing great flexibility to attackers. We have also implemented two other covert attacks with the portable device and everyday life loudspeaker, as shown in Fig.~\ref{fig:portable_device}.
\subsubsection{Portable Device}
\begin{figure}[t]
	\centering  
        \subfigure[Portable Device]{   
    	\begin{minipage}[t]{0.23\textwidth}
    		\centering
    		\includegraphics[width=1.06\textwidth]{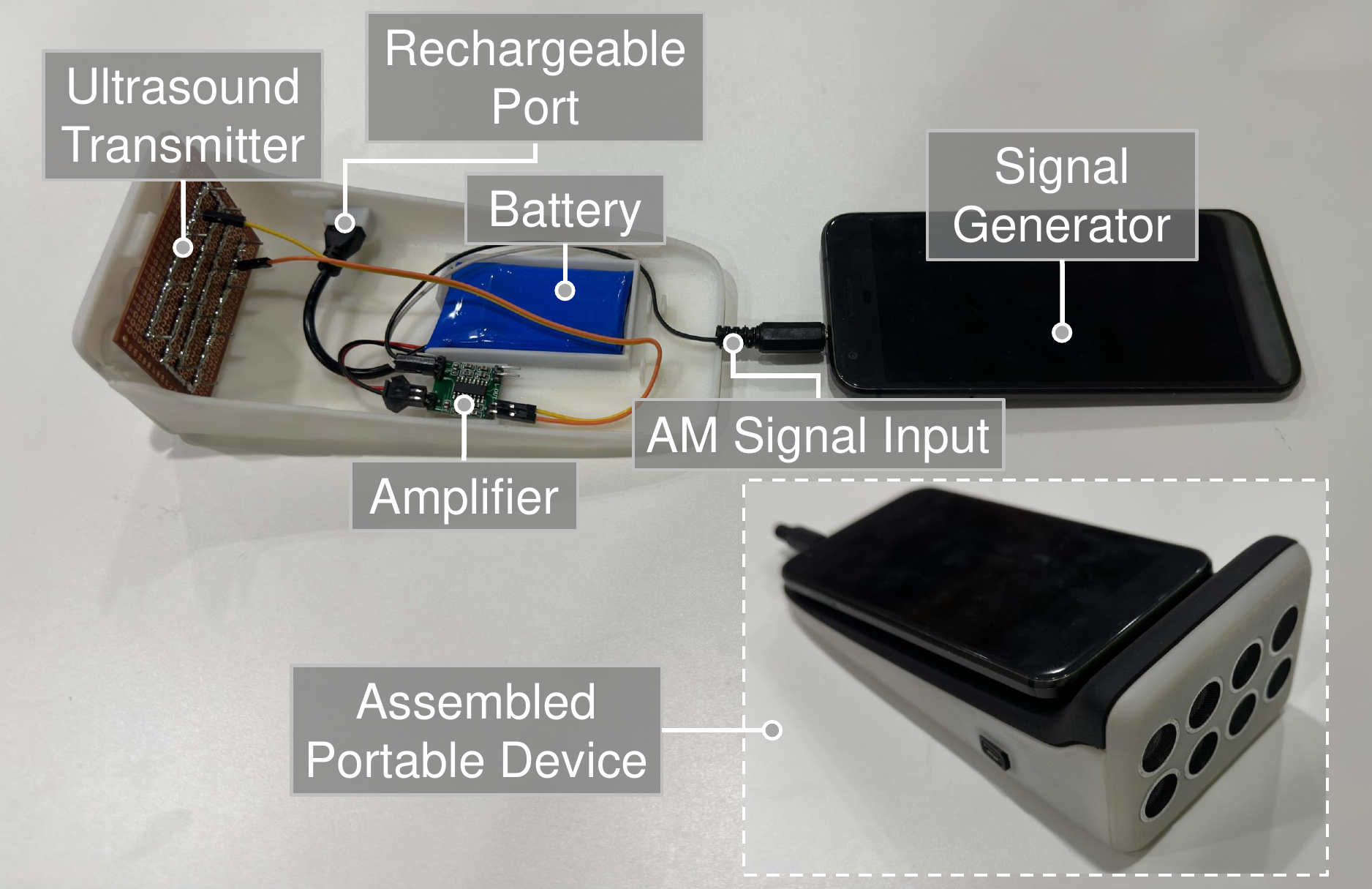} 
    	\end{minipage}
        }\hfill
	\subfigure[Off-the-shelf Loudspeaker]{   
	\begin{minipage}[t]{0.22\textwidth}
		\centering
		\includegraphics[width=1.03\textwidth]{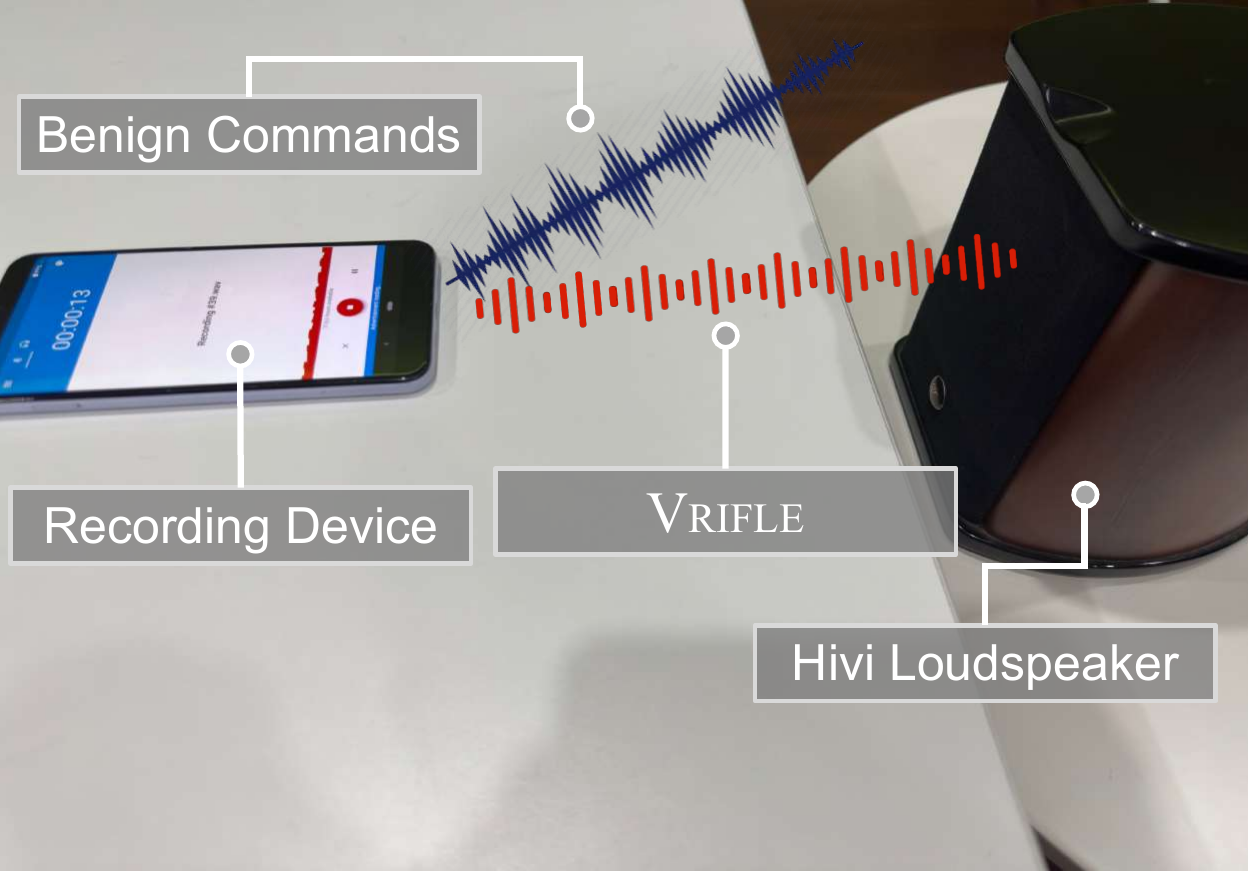} 
	\end{minipage}
	}
	\caption{Two additional attack forms of \alias.}    
	\label{fig:portable_device}    
	\vspace{-10pt}
\end{figure}

Our portable device equipped with eight 25~kHz ultrasound transducers, a compact amplifier, and a rechargeable battery in Fig.~\ref{fig:portable_device}(a), balances lightweight and attack range. It can be connected to the smartphone, where the attacker stores perturbations as 96~kHz USB-AM audio in advance. We evaluate the effectiveness of attacks with a portable device, setting it to point at Mix2s' bottom microphone with the target ``open the door''. Fig.~\ref{fig:eval_portable_hivi}(a) demonstrates 100\% SR within 150~cm, and 78\% SR along with CER down to 1.69\% even at a distance of 180~cm, suggesting \alias with portable devices can exceed the attack distance of almost prior AEs.

\subsubsection{Off-the-shelf Loudspeaker}
\begin{figure}[t]
	\centering  
        \subfigure[Portable Device]{   
    	\begin{minipage}[t]{0.23\textwidth}
    		\centering
    		\includegraphics[width=1\textwidth]{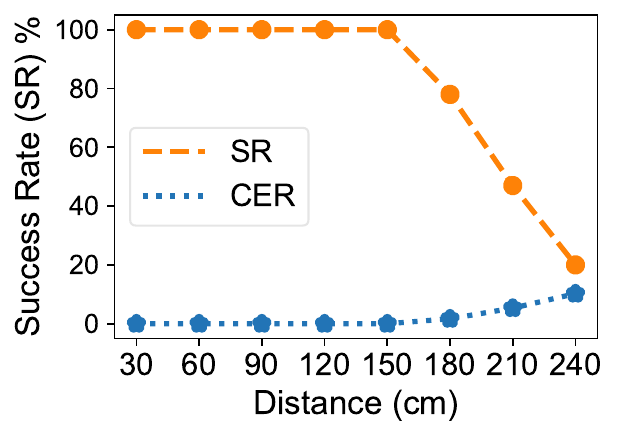} 
    	\end{minipage}
        }\hfill
	\subfigure[Off-the-shelf Loudspeaker]{   
	\begin{minipage}[t]{0.207\textwidth}
		\centering
		\includegraphics[width=1\textwidth]{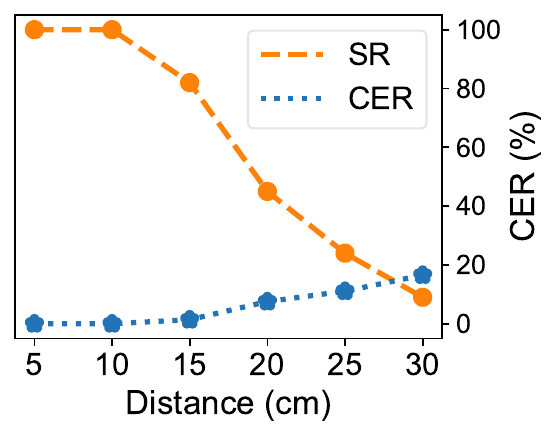} 
	\end{minipage}
	}
	\caption{\alias's performance at different distance with the portable device and off-the-shelf loudspeaker.}    
	\label{fig:eval_portable_hivi}    
	\vspace{-10pt}
\end{figure}

Adversaries can embed USB-AM perturbations into audio or video files to manipulate user commands when played on a computer or smartphone connected to a loudspeaker. We investigate the use of off-the-shelf loudspeakers, such as the high-end Hivi~\cite{hivi}, which have three distinct sound sources: woofer (37-140~Hz), mid-range (140-2000~Hz), and tweeter ($>$2000 Hz).
To determine the optimal ultrasound frequency for embedding the perturbations, we conduct experiments scanning the carrier frequency from 21-27~kHz and find 25.2~kHz to be the best frequency, despite the gain decrease beyond the rated frequency range (37Hz$\sim$20kHz).
Figure\ref{fig:eval_portable_hivi}(b) illustrates that \alias's effective attack distance via off-the-shelf speakers is approximately 20~cm, with a low CER of 11.07\%, demonstrating effective modification of user commands at the character level. 

\section{Anti-defense Experiment}
In this section, we validate whether \alias can resist 6 kinds of representative defenses, involving audio pre-processing methods and inaudible attack detection. We consider two types of adversaries: 1) \emph{Naive Adversary}: The naive adversary creates \alias based on the undefended model to attack the defended model. 2) \emph{Adaptive Adversary}: This adversary has full knowledge of the defense mechanisms and applies customized strategies to craft \alias.

\begin{table}[t]
\footnotesize
\centering
\renewcommand\arraystretch{0.8} 
\setlength\tabcolsep{2.8pt}
\caption{Three defenses}
\begin{tabular}{c|c|c|c|c}
\toprule[1pt]
\multicolumn{1}{c|}{\textbf{(\%) Method}} & \multicolumn{1}{c|}{\textbf{Undefended}} & \textbf{Quantization} & \textbf{VAD}   & \textbf{Opus Compress} \\ \midrule[0.8pt]
\multicolumn{1}{c|}{Success Rate}                      & \textbf{99.49}                           & 97.96        & 99.49 & 95.93         \\ \midrule
\multicolumn{1}{c|}{Attack CER}                      & \textbf{0.1}                            & 0.28         & 0.1  & 0.84          \\ \midrule
\multicolumn{1}{c|}{Benign CER }                     & \textbf{11.23}                           & 13.43        & 23.58 & 12.37         \\ \bottomrule[1pt]
\end{tabular}
\label{tab:definitive}
\vspace{-5pt}
\end{table}

\begin{table}[t]
\footnotesize
\centering
\renewcommand\arraystretch{0.8} 
\setlength\tabcolsep{1pt}
\caption{Defense with band-pass filter}
\begin{tabular}{cc|c|c|c|c|c}
\toprule[1pt]
\multicolumn{2}{c|}{\textbf{(\%) Band-pass (Hz)}}                                                                               & \multicolumn{1}{l|}{\textbf{50$\sim$7000}} & \multicolumn{1}{c|}{\textbf{50$\sim$6000}} & \multicolumn{1}{l|}{\textbf{50$\sim$5000}} & \multicolumn{1}{l|}{\textbf{50$\sim$4000}} & \multicolumn{1}{l}{\textbf{50$\sim$3000}} \\ \midrule[0.8pt]
\multicolumn{1}{c|}{\multirow{2}{*}{\multirow{2}{*}{\begin{tabular}[c]{@{}c@{}}\emph{Naive}\\ \emph{Adversary}\end{tabular}}}}    & Success Rate & 97.96                             & 96.95                             & 95.67                             & 75.57                             & 12.47                            \\ \cmidrule{2-7} 
\multicolumn{1}{c|}{}                                                                              & Attack CER   & 0.28                              & 0.53                              & 0.80                              & 6.43                             & 35.01                            \\ \midrule[0.8pt]
\multicolumn{1}{c|}{\multirow{2}{*}{\multirow{2}{*}{\begin{tabular}[c]{@{}c@{}}\emph{Adaptive}\\ 
\emph{Adversary}\end{tabular}}}} & Success Rate & 99.75                             & 99.24                            & 97.20                             & 91.60                             & 78.88                            \\ \cmidrule{2-7} 
\multicolumn{1}{c|}{}                                                                              & Attack CER   & 0.05                              & 0.14                              & 0.41                              & 1.64                              & 5.34                            \\ \midrule[0.8pt]
\multicolumn{2}{c|}{Benign CER}                                                                                   & 11.63                             & 13.60                             & 19.27                             & 28.83                             & 41.06                            \\ \bottomrule[1pt]
\end{tabular}
\label{tab:bandpass}
\vspace{-5pt}
\end{table}

\begin{table}[t]
\footnotesize
\centering
\renewcommand\arraystretch{0.8} 
\setlength\tabcolsep{2.8pt}
\caption{Defense with down-sampling}
\begin{tabular}{cc|c|c|c|c|c|c}
\toprule[1pt]
\multicolumn{2}{c|}{\textbf{(\%) Down-sample (rate)}}                                                                               & \multicolumn{1}{c|}{\textbf{0.9}} & \multicolumn{1}{c|}{\textbf{0.8}} & \multicolumn{1}{c|}{\textbf{0.7}} & \multicolumn{1}{c|}{\textbf{0.6}} & \multicolumn{1}{c|}{\textbf{0.5}} & \multicolumn{1}{c}{\textbf{0.4}}  \\ \midrule[0.8pt]
\multicolumn{1}{c|}{\multirow{2}{*}{\multirow{2}{*}{\begin{tabular}[c]{@{}c@{}}\emph{Naive}\\ \emph{Adversary}\end{tabular}}}}    & Success Rate & 98.98                             & 98.22                             & 94.91                             & 88.04                             & 69.47  & 19.34                            \\ \cmidrule{2-8} 
\multicolumn{1}{c|}{}                                                                              & Attack CER   & 0.20                              & 0.27                             & 1.00                              & 2.58                              & 8.35    &  30.23                            \\ \midrule[0.8pt]
\multicolumn{1}{c|}{\multirow{2}{*}{\multirow{2}{*}{\begin{tabular}[c]{@{}c@{}}\emph{Adaptive}\\ 
\emph{Adversary}\end{tabular}}}} & Success Rate & 99.24                             & 98.22                             & 96.69                             & 95.42                             & 89.06  & 81.42                            \\ \cmidrule{2-8} 
\multicolumn{1}{c|}{}                                                                              & Attack CER   & 0.14                              & 0.27                             & 0.57                              & 0.87                              & 2.25    &  4.38                            \\ \midrule[0.8pt]
\multicolumn{2}{c|}{Benign CER}                                                                                   & 11.51                             & 12.11                             & 14.93                             & 19.80                             & 23.99   &  35.04                            \\ \bottomrule[1pt]
\end{tabular}
\label{tab:squeezing}
\end{table}

\textbf{Against Audio Pre-processing Defense Methods.} 
Referring to previous works~\cite{hussain2021waveguard,deng2022fencesitter,li2020advpulse,yuan2018commandersong} that present a series of audio pre-processing methods against audio adversarial example attacks, we examine the robustness of \alias using 5 representative defenses: \textit{(1) Quantization}: converting the audio sampling value from a 16-bit signed integer to an 8-bit precision, which reduces the sampling range from [-32,768$\sim$32,767] to [-128$\sim$127]. Notably, this introduces distortion and noise due to the small range of values at 8-bit precision. \textit{(2) Voice Activity Detection (VAD)}: removing segments of audio that are less than -15~dB, where its maximum energy is normalized to 0~dB. \textit{(3) Opus Compression Codec}: coding and compressing audio with flexible bit rate and low latency are widely used in real-time communication, particularly VoIP and online meetings. We set the default compression level as 5 according to \cite{torch_codec}. \textit{(4) Band-pass Filter}: filtering the input signal with given cut-off frequencies, e.g., 50$\sim$7000 Hz. \textit{(5) Down-sampling}: reducing the audio to a given rate, e.g., rate=0.4 means down-sampling a 16 kHz audio to 6.4 kHz, and then recovering it to the required sampling rate of targeted ASRs (generally 16/48~kHz).

We obtain the attack success rate (successfully altering the tested speech into the targeted command ``open the door''), attack CER, and benign CER (derived between the DeepSpeech recognized and ground-truth transcription) at 99.49\%, 0.15\%, 11.23\%, respectively, when the model is undefended.
The results are listed in Tab.~\ref{tab:definitive}, \ref{tab:bandpass}, \ref{tab:squeezing}. We observe that quantization, VAD, and Opus compression barely affect the attack success rate (all$\ge$95.93\%) in Tab.~\ref{tab:definitive}. Particularly, VAD significantly rises benign CER from 11.23\% to 23.58\%, while failing to lower our attack performance. Tab.~\ref{tab:bandpass} and \ref{tab:squeezing} demonstrate that naive \alias can maintain relatively effective even when facing a 50$\sim$4000 band-pass filter or being down-sampled to 8 kHz (rate=0.5). However, the attack performance degrades as the bandwidth or the down-sampling rate gets further lower. Note that we do not evaluate extreme cases, such as band-pass: less than 50$\sim$2000 Hz or down-sampling rate: smaller than 0.3, since they have severely affected the model's ability to transcribe benign speech commands with unacceptable CERs over 45\%. After the adaptive adversary integrates the band-pass filtering operation during optimization, the attack performance increase significantly, especially the success rate and attack CER reach 78.88\% and 5.34\%, respectively, even under 50$\sim$3000 Hz band-pass filtering. Similarly, the adaptive adversary can realize an 81.42\% success rate and 4.38\% CER against a down-sampling rate=0.4.



\begin{figure}[t]
    \centering
    \includegraphics[width=0.34\textwidth]{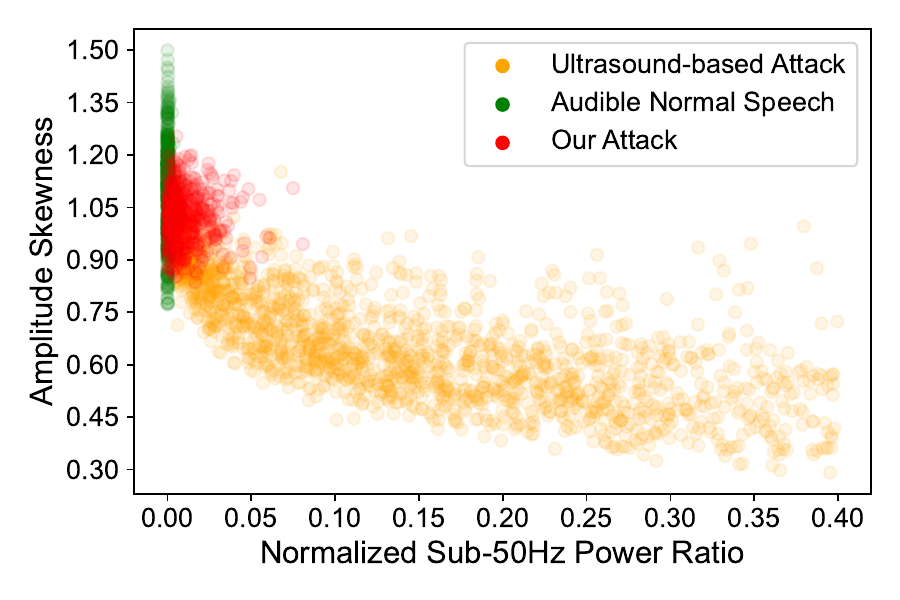}
    \caption{Two significant feature dimensions extracted from three classes of audio samples by LipRead (800 samples/class).}
    \label{fig:detect_compare}
    \vspace{-10pt}
\end{figure}

\textbf{Against Inaudible Attack Detection Method.} 
Given that \alias utilizes ultrasound-based modulation mechanisms, prior inaudible attack detection methods are expected to distinguish such an attack well from benign speech. We reproduce the representative software-based method: LipRead~\cite{roy2018inaudible}, strictly following its instruction, which extracts and analyzes three features of speech samples: power in sub-50Hz, correlation coefficient (between the fundamental and harmonic components), and amplitude skew. 
We use the LipRead classifier to detect \alias samples crafted under the naive adversary setting and collected at different distances \& angles; then obtain a detection accuracy down to 45.07\%. Fig.~\ref{fig:detect_compare} visualizes three types of audio samples in two significant feature dimensions. \alias presents compact skewness around 1.0 due to its symmetrical waveform, whose distribution is closer to the normal, while ultrasound-based attacks appear more shift toward 0.30 and greater power in sub-50Hz. Low-frequency power aggregation is still inevitable in our attack due to nonlinear demodulation~\cite{roy2018inaudible}. Moreover, naive \alias appears low correlation coefficient compared to the traditional attacks, as its perturbations (see Fig.~\ref{fig:design_overview}\&\ref{fig:design_attack}) barely present normal speech properties such as fundamental and harmonic frequencies. 
Overall, the inherent difference between \alias and traditional ultrasound-based attacks makes it probably compromise LipRead.
Furthermore, the adaptive adversary extracts three features during the perturbation generation and constrains them close to the normal samples, further reducing the accuracy of LipRead detecting our attack to 30.55\%.
\vspace{-5pt}
\section{Discussion and Future Work}\label{sec:discuss}
%
\hspace{0.3em}\textbf{Potential Countermeasure:}
We have demonstrated that \alias are robust to audio pre-processing and inaudible attack detection methods. We envisage that defense approaches tracking ultrasound nature~\cite{zhang2021eararray,he2019canceling} may be effective, although these methods are based on two hardware-dependent prototypes that can not adapt to off-the-shelf compact smart devices. For the remaining feature forensics-based~\cite{zhang2017dolphinattack,roy2018inaudible} or ML-based defenses~\cite{li2021robust,li2023learning}, we believe that the adaptive adversary shall adopt these defense strategies along with the ultrasonic transformation model during optimization and physically bypass them. But this would result in a less universal attack due to additional constraints.

\textbf{Prevent Airborne Self-demodulation Leakage.}
Although our attack distance has significantly exceeded previous works, please note that \alias cannot extend the range infinitely. As uncovered in \cite{iijima2018audio}, the self-demodulation occurs and then the modulated baseband becomes audible once a certain power is reached. To increase the attack range while ensuring inaudibility, we adopt the following strategies: 1) utilizing 25 kHz carrier frequency rather than higher frequencies, such as 40kHz, for less attenuation; 2) employing customized ultrasound transducers, signal generator, and amplifiers capable of suppressing nonlinear distortion at the speaker side; and 3) setting maximum power not to exceed 3.2W and implementing USB-AM to increase the attack efficiency with portable device and off-the-shelf loudspeakers. 

\textbf{Limitations:}
1) \alias achieves highly universal manipulation of user speech using DeepSpeech2's gradient information. However, its universality under black-box settings is limited in critical user-present scenarios due to variable user factors. 
Notably, targeted universal AE attacks in black-box scenarios remain an unsolved problem currently, despite several untargeted literature~\cite{vadillo2019universal,neekhara2019universal}.
2) Although we have verified that our ultrasonic transformation model is effective on different recording devices, it is currently device-specific due to the microphone's frequency selectivity to ultrasound. We will investigate a device-generic transformation model in future work. 
3) Our careful design enables the \textit{man-in-the-middle} attack strategy and our user testing in Appendix \textsection\ref{append:user_test} demonstrate its high stealthiness. However, the testing results imply that replaying excessively long user commands may cause discomfort and might alert the user. We envision that understanding user intent and then replaying synthetic short commands can mitigate this issue.


\textbf{Attack on Speaker Recognition:} 
We envision that the idea of \alias can be generalized to attack speaker recognition models deployed on access control systems, e.g., authentication of voice assistants and applications. We have conducted a preliminary experiment attacking the state-of-the-art ECAPA-TDNN~\cite{2020ecapatdnn}, a popular speaker recognition model. We maintain the almost identical design as used in attacking the ASR model and only reconfigure the optimization goal $y_t$ as the target speaker label and the loss function $\mathcal{L}(f(\cdot),y_t)$ as the cosine similarity scoring module. Results demonstrate that, in a 10-person set, \alias is universal to alter the voiceprint of any user's speech samples into the targeted speaker's. We plan to delve into such an ability of \alias in future work.

\section{Related Work}
\textbf{Custom Adversarial Examples \& Inaudible Attacks.}
The initial AE attacks construct a custom (i.e., non-universal) perturbation for a specific audio clip, whereas the same perturbation cannot compromise other audio. Signal-level transformations~\cite{vaidya2015cocaine,carlini2016hidden,abdullah2019ndss}, such as modifying MFCC, are unintelligible to human beings but can be recognized by the ASR model. As this class of attacks resembles obvious noises, they can easily alert users. Thus, inaudible attacks~\cite{zhang2017dolphinattack,roy2018inaudible,yan2020surfingattack} have been proposed, which exploit carrier signals outside the audible frequencies of human beings (e.g., 40~kHz) to inject attacks into ASR systems utilizing the nonlinearity vulnerability of microphone circuits, yet entirely unheard by victims. However, compared with audible playback speech samples, such attacks usually suffer from signal distortion and low SNR due to their dependence on various convert channels, e.g., ultrasound~\cite{ji2022capspeaker}, laser~\cite{sugawara2020light}, or electricity~\cite{wang2022ghosttalk} signals, and the hardware imperfections these channels introduce.
There is also a major branch of the research community that leverages the vulnerability of ASR models by adding slightly audible perturbations on the benign audio based on $\epsilon$-constraint~\cite{carlini2018audio,taori2019targeted} and psychoacoustic hiding~\cite{schonherr2018adversarial,qin2019imperceptible}, to make the AEs sound benign but fool the ASR's transcription. It is worth noting that non-universal AEs lose effectiveness for streaming speech input and unpredictable user commands, as they rely on perfect temporal alignment. Constructing multiple AEs for altering different commands as an adversary-desired instruction is also impractical. 

\textbf{Universal Adversarial Examples.}
Recent studies propose universal AEs that can apply to tamper with multiple speech content as an adversary-desired command. Existing untargeted universal AE attacks adopt iterative greedy algorithms~\cite{moosavi2017universal} can cause arbitrary speech to mis-classification~\cite{vadillo2019universal} or false transcription~\cite{neekhara2019universal}. In contrast, targeted universal AE attack is very challenging in speech recognition tasks because ASR models are context-dependent, and a certain minor perturbation superimposed even at different positions of a given benign audio, the whole sentence may yield various transcription results. This is distinct from the prior successful targeted universal AE attack in the text-independent speaker recognition~\cite{deng2022fencesitter,li2023tuner} and the universal adversarial patch attack in position-insensitive CNN-based image classification tasks~\cite{brown2017adversarial}. Moreover, given that the victim user can easily notice the audible-band perturbation, AdvPulse~\cite{li2020advpulse} disguises short pulses in the environment sounds to be less perceptible. However, they only apply to a context-insensitive CNN-based audio command classification model to be universal. 
\blue{
To overcome the mainstream RNN-based ASR context-dependent issues, a partial match strategy is proposed by SpecPatch~\cite{guo2022specpatch}, which also employs audible noise-like short pulses (0.5s) to alter multiple short user commands into the targeted instruction against the mainstream DeepSpeech ASR model. However, such an attack will not work in relatively long commands ($\ge$ 4 words) and can be noticed despite following L2-imperceptibility constraints.}

\blue{
Overall, due to the fundamental differences between audible and ultrasonic channels, \alias differs from prior works that encountered challenges related to \textbf{\textit{user auditory}} and \textbf{\textit{user disruption}}. 
In addition to the four representative merits over existing AEs listed in Tab.~\ref{tab:compare}, \alias offers several additional benefits: (1) the optimization process is no longer subject to audibility constraints such as tiny $\epsilon$, psychoacoustics, $L_p$-norm, nor does it need to limit the signal form as short pulses to reduce the possibility of being perceived. (2) \alias's broad optimization space further allows for fewer iterations while maintaining a high degree of universality. Combining these two advantages, \alias enables real-time manipulation of arbitrary user commands and long speech sentences in an \textit{alter-and-mute} fashion, as never before. (3) Unlike audible-band AEs that are easily compromised by interference due to their subtle perturbations, \alias demonstrates robustness and remains effective even when faced with various audio pre-processing defenses.
Notably, our initial modeling of ultrasound transformation precisely characterizes the ultrasound channel and justifies it as a promising carrier for IAP delivery. We believe that this modeling effort lays the groundwork for generating inaudible AEs and may inspire future works.
}

\color{black}

\section{Conclusion}

In this work, we propose an inaudible adversarial perturbation (IAP) attack against ASR systems named \alias, which can extend to scenarios where users are present and may use ASR services. In such scenarios, prior studies will fail due to \textit{user auditory} and \textit{user disruption}. 
We make the first attempt to model the ultrasonic transformation process, based on which, \alias can alter arbitrary user commands to the adversary-desired intent in real time without any knowledge of users' speech. Our comprehensive experiments in the digital and physical worlds across various configurations demonstrate \alias's effectiveness and robustness. Overall, \alias features merits including complete inaudibility, universality, and long-range attack ability.
\section*{Acknowledgement}
We thank the shepherd and anonymous reviewers for their valuable comments and suggestions. We also appreciate Dr. Jiangyi Deng for providing valuable insights and guidance to the paper's content. This work is supported by China NSFC Grant 61925109, 62201503, 62222114, and 62071428.


%
\bibliographystyle{unsrt} 
\bibliography{refs}
~
\section*{Appendix}
\setcounter{equation}{0}
\setcounter{subsection}{0}
\setcounter{section}{0}
\renewcommand{\theequation}{\arabic{equation}}
\renewcommand{\thesubsection}{\arabic{subsection}}
\renewcommand{\thesection}{\Alph{section}}

\section{Single-sideband Amplitude Modulation}\label{append:SSB}
In this section, we give mathematical proof that the baseband perturbation of SSB-AM signals can be recovered by commercial microphones. We initially compare the maximum energy of USB-AM and LSB-AM emitting the same perturbation when sound leakage occurs, and LSB-AM is 87\% of USB-AM. Thus, we adopt the USB-AM in our attacks due to its better inaudibility:
\begin{equation*}
\begin{aligned}
    & \text{USB-AM:}~ S_{USB}(t)=m{cos\omega_{c}t}-\hat{m}{sin\omega_{c}t}+cos\omega_{c}t\\
    & \text{LSB-AM:}~ S_{LSB}(t)=m{cos\omega_{c}t}+\hat{m}{sin\omega_{c}t}+cos\omega_{c}t
\end{aligned}
\label{equ:SSB_formula}
\end{equation*}
where the $\hat{m}$ is the conjugate of $m$. The microphone amplifier's output is below:
$$S_{out} = k_{1}S_{USB}(t) + k_{2}S_{USB}^2(t) + \cdots$$
The $S_{USB}^2(t)$ term has three components: a high-frequency $2\omega_{c}t$ components:
$$(m+1) \hat{m} \sin(2 \omega_c t)+\frac{m^2+2m+1-\hat{m}^2}{2} \cos(2\omega_c t)$$
a direct current (DC) term $\frac{1}{2}$ and an audible component $S_{aud}(t)=\frac{1}{2}(m^2+2m+\hat{m}^2)$.
$S_{USB}(t)$ and the high-frequency component are filtered by the low-pass filter because its frequency is above 25~kHz. The DC component is filtered by the microphone’s capacitor. Thus, the audible component $S_{aud}(t)$ that passes the microphone filtering system can function to ASR.

\section{Real-world Scenario}\label{append:attack_scenario}
\begin{figure}[h]
	\centering
	\includegraphics[width=0.48\textwidth]{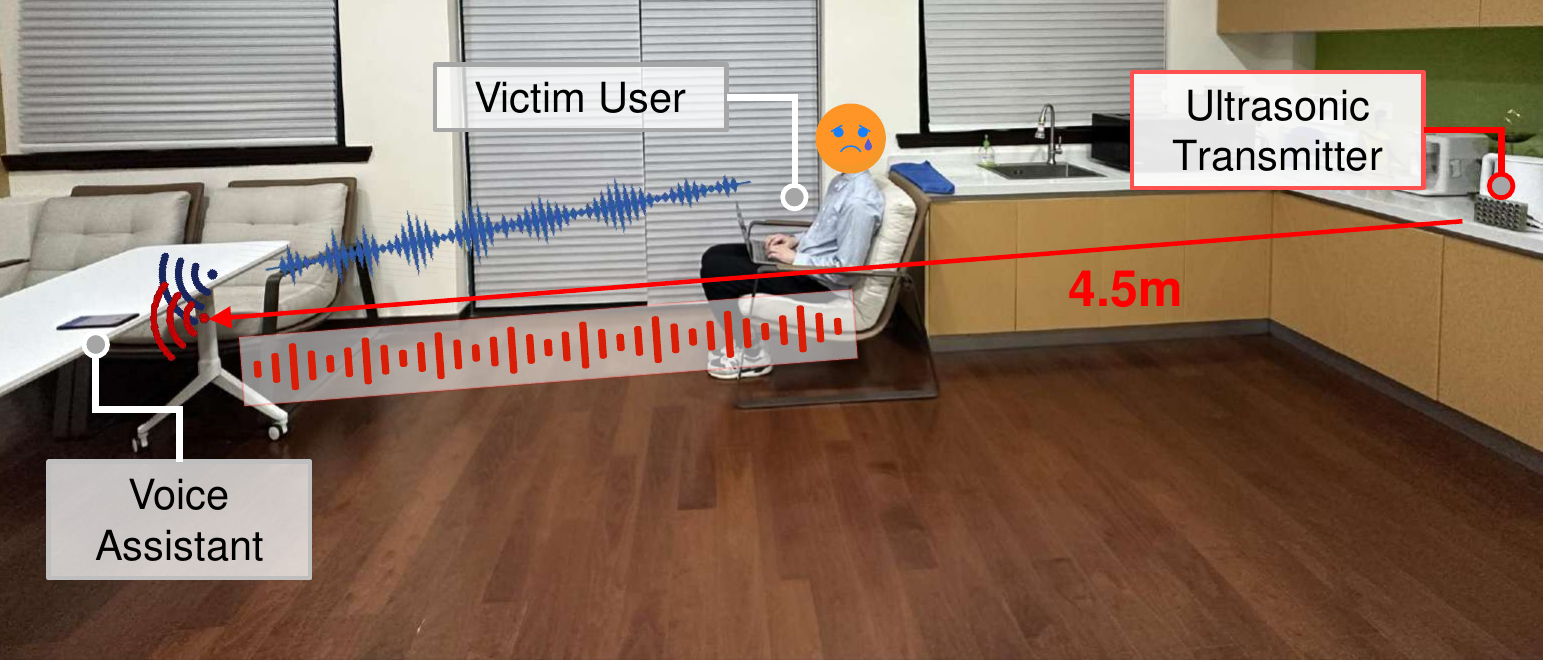}
	\caption{\label{fig:attack_actual}Real-World Attack Scenario.}
\end{figure}

Figure~\ref{fig:attack_actual} presents our real-world attack scenario.

\section{Targeted Commands Lists}\label{append:command_list}
Tab.\ref{tab:diff_commands} lists 10 different commands, corresponding to the performance of constructing target command-specific perturbations in experiment \textsection\ref{sec:eval_commands}.
\begin{table}[h]
	\centering
        \normalsize
		\caption{Attack with Different Targeted Commands}
		\renewcommand\arraystretch{0.9}
		\renewcommand\tabcolsep{1.5pt}
		\begin{threeparttable}
			\begin{tabular}{l|c|c}
                \toprule
                \textbf{Target Command} & \textbf{~~~~SR~~~~} & \textbf{~~CER~~} \\
				\midrule
                    ``Start recording''  & 100\% & 0\% \\ \midrule
				``Set a timer''  & 100\%  & 0\% \\ \midrule
				``Open the door''  & 100\% & 0\% \\ \midrule
				``Take the picture''  & 100\% & 0\% \\ \midrule
				``Call nine one one (911)''  & 100\% & 0\% \\ \midrule
				``Cancel my morning alarm''  & 100\% & 0\% \\ \midrule
                    ``Turn on airplane mode'' & 94.39\% & 0.28\% \\ \midrule
                    ``Open my photo album''  & 95.03\% & 0.50\% \\ \midrule
				``What is going on Twitter?''  & 100\% & 0\% \\ \midrule
                    ``Mute volume and turn off the WiFi'' & 92.82\% & 0.21\% \\
				\bottomrule
			\end{tabular}
		\end{threeparttable}
		\label{tab:diff_commands}
\end{table}
\vfill\eject

\section{Algorithm of \alias}\label{append:algo1}
Given that technical workflow for the silence and universal perturbation are overall identical, the major differences are the optimization objective: $y_t/y_b$ and hyper-parameters. Therefore, we demonstrate \alias's representative optimization process of crafting a universal perturbation from scratch in Algorithm.~\ref{algo1}.

\begin{algorithm}[h]
	\caption{Universal \alias Generation}
	\label{algo1}
	\LinesNumbered
	\KwIn{The ASR model with CTC Loss Computation module: $\mathcal{L}$, the maximum epoch: $\text{maxEpoch}$, the desired loss: $objValue$, with a scoring module: $S$, the learning rate: $\eta$, the preset time range: $T$.}
	\KwOut{The universal perturbation $\delta$}
	\textbf{Init} $\delta \gets 0^N$\\
	\For{$1$ to ${ maxEpoch}$}
	{
		${J} \gets 0$\\
            \For{$h_{\theta}\in U_H$, $n\in U_N$}
            {
                $\hat{e} = e^{-a_0 \omega_{c}^{n}d}$\\
                $\overline{\delta} = h_{\theta}\hat{e}\ast \overline{\delta:\hat{\xi}} + n$\\
                \For{$x\in U_x, g\in G, S_{(\cdot)}~{s.t.}~T$}
                {
                    $\Tilde{x}=\beta\cdot g\ast x$\\
                    $\Tilde{x_{\delta}} = clip(\Tilde{x}+\mathcal{S}_{(\overline{\delta})}, [-1,1])$\\
                    $J+=\mathcal{L}(\Tilde{x_{\delta}}, y_t)$
                }
            }
            Compute ${\nabla}_\mathcal{\delta}J$\\
            $\delta \gets \Omega_{Adam}(\delta+\eta\cdot {\nabla}_\mathcal{\delta}J )$\\
            $\delta \gets clip(\delta, [-1,1])$\\
            \If{$J \le objValue$}{break}
        }
	\normalsize
\end{algorithm}


\section{Different Speech \& Perturbation Loudness}\label{append:loudness_compare}
\begin{figure}[h]
	\centering  
		\subfigure[Success Rate (\%)]{   
		\begin{minipage}[t]{0.22\textwidth}
			\centering
			\includegraphics[width=1.1\textwidth]{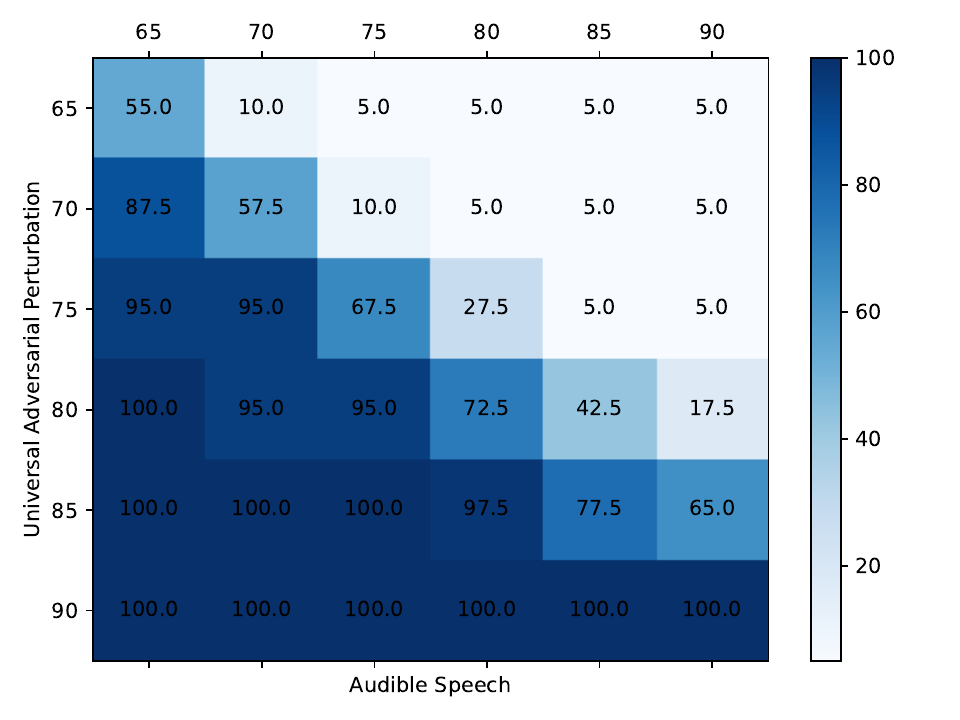} 
		\end{minipage}
		}
	\subfigure[CER (\%)]{   
	\begin{minipage}[t]{0.22\textwidth}
		\centering
		\includegraphics[width=1.1\textwidth]{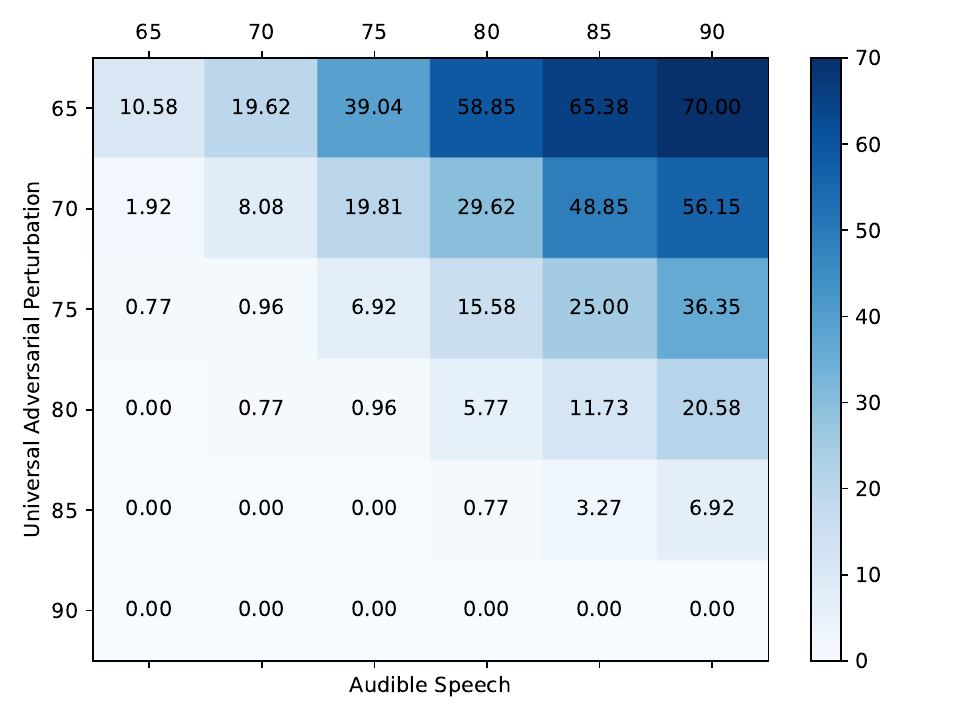} 
	\end{minipage}
	}
	\caption{The performance of loudness relationship between user speech and perturbation.}    
	\label{fig:loudness_compare}    
\end{figure}

Fig.~\ref{fig:loudness_compare} shows the success rate and CER of our experiments on the relative energies between the attack perturbation and speech.

\section{User Testing}\label{append:user_test}
\begin{figure}[h]
    \centering
    \includegraphics[width=0.45\textwidth]{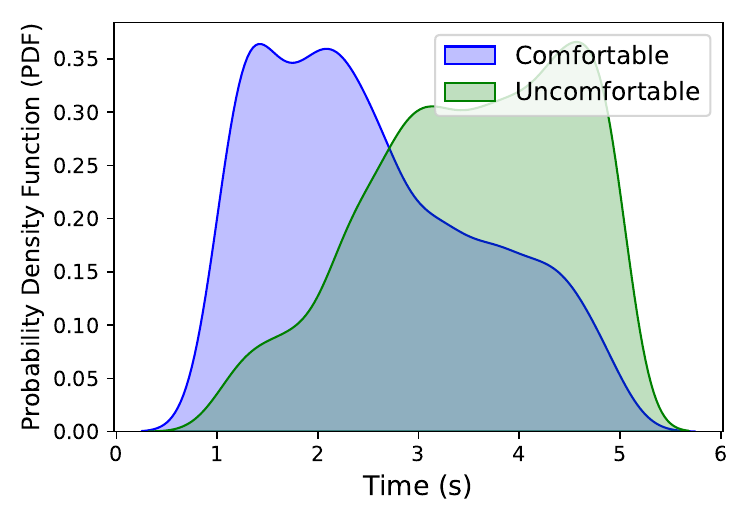}
    \caption{The probability distribution of users' awareness during a \textit{man-in-the-middle} attack under different delay conditions (similar to network latency). ``Comfortable'': the situation where users find the ASR service is normal and are not aware of the attack; ``Uncomfortable'': the delay may cause them to feel uncomfortable or unusual.}
    \label{fig:user_latency}
\end{figure}

In this section, we elaborate on the \textit{Man-in-the-middle} attack strategy, whose effect is akin to experiencing network congestion when users use the ASR service, resulting in slower responses. Prolonged latency can make users feel uncomfortable while using the service. 
To assess user awareness under such delays, we design 10 scenarios, each consisting of an audio clip that simulates a user issuing a command to the ASR system with random delays (1-5 seconds) before the voice assistant executes the command. 
We collected test results from 140 college students of different majors. As shown in Figure~\ref{fig:user_latency}, when the delay time is less than 2.7 seconds (the junction point of two distribution curves), more users find the ASR service comforting than uncomfortable. 
The participants are also asked to fill in what they think the cause is if they experience an uncomfortable delay when using the ASR service.
Only 11 out of 140 participants suspect an attack, while almost all others attribute the delay to network latency/congestion or device stuck, suggesting that this strategy poses a hidden attack.
We believe that users' suspicion may also be related to their disciplinary background, e.g., users with knowledge of cybersecurity are more likely to consider the possibility of an attack.

\end{document}